\documentclass[showpacs,preprintnumbers,
superscriptaddress,amsmath,floatfix]{revtex4}

\usepackage{graphicx}

\newcommand{\PLB}[3]{Phys.\ Lett.\ B\ {\bf #1},\ #2 (#3)}

\newcommand{\PRL}[3]{Phys.\ Rev.\ Lett.\ {\bf #1},\ #2 (#3)}

\newcommand{\PRD}[3]{Phys.\ Rev.\ D\ {\bf #1},\ #2 (#3)}

\newcommand{\ibid}[3]{{\bf #1},\ #2 (#3)}

\renewcommand\a{\alpha}
\renewcommand\b{\beta}
\newcommand\g{\gamma}
\renewcommand\d{\delta}
\newcommand\e{\epsilon}

\newcommand\m{\mu}
\newcommand\n{\nu}

\newcommand\D{\Delta}
\newcommand\G{\Gamma}

\newcommand{\be}{\begin{equation}}
\newcommand{\ee}{\end{equation}}
\newcommand{\bea}{\begin{eqnarray}}
\newcommand{\eea}{\end{eqnarray}}
\newcommand{\ba}[1]{\begin{array}{#1}}
\newcommand{\ea}{\end{array}}

\newcommand{\bm}[1]{\mbox{\boldmath${#1}$}}
\newcommand{\uq}{\hat{\mathbf{q}}} 
\newcommand{\uk}{\hat{\mathbf{k}}}
\newcommand{\up}{\hat{\mathbf{p}}}
\newcommand{\hk}{\hat{k}}
\newcommand{\hq}{\hat{q}}
\newcommand{\hp}{\hat{p}}
\newcommand{\vg}{\bm{\gamma}}
\newcommand{\gperp}{\bm{\gamma}_{\perp}}

\newcommand{\vv}{\bm{v}}
\newcommand{\vJ}{\bm{J}}

\newcommand{\Tr}{{\rm Tr}}
\newcommand{\angk}[1]{\langle #1 \rangle_{\uk}} 
\newcommand{\angq}[1]{\langle #1 \rangle_{\uq}}

\begin{document}

\title{The ground state in a spin-one color superconductor} 

\author{Andreas Schmitt}
\email{aschmitt@th.physik.uni-frankfurt.de}
\affiliation{Institut f\"ur Theoretische Physik, 
J.W. Goethe-Universit\"at, D-60054 Frankfurt/Main, Germany}
\affiliation{New address from January 1st, 2005: Center for Theoretical Physics,
MIT, Cambridge, MA 02139}

\date{\today}

\begin{abstract}

Color superconductors in which quarks of the same flavor form Cooper pairs 
are investigated. These Cooper pairs carry total spin one.
A systematic group-theoretical classification of possible 
phases in a spin-one color superconductor is presented, revealing parallels
and differences to the theory of superfluid $^3$He.
General expressions for the gap parameter, the critical temperature,  
and the pressure are derived and evaluated for several spin-one phases, with  
special emphasis on the angular structure of the gap equation. 
It is shown that the (transverse) color-spin-locked
phase is expected to be the ground state.

\end{abstract}
\pacs{12.38.Mh,24.85.+p}

\maketitle

\section{Introduction} \label{intro}

It is well-known that certain metals and alloys exhibit a superconducting
phase below a critical temperature $T_c$. In this phase, electrons in the
vicinity of the Fermi surface form Cooper pairs which leads to a gap 
$\phi$ in the quasiparticle excitation spectrum \cite{bcs}. An attractive 
interaction between the electrons is provided by the exchange of virtual 
phonons, and the electromagnetic gauge group $U(1)_{em}$ is 
spontaneously broken. A similar mechanism occurs in sufficiently cold
and dense quark matter \cite{bailin}. In this case, the attractive 
color-antitriplet
channel is responsible for the formation of quark Cooper pairs. Due to the 
intrinsic properties of quarks (color, flavor, electric charge), many 
pairing patterns seem to be theoretically possible. In other words, besides
the electromagnetic gauge group, also the color gauge group $SU(3)_c$,  
the flavor group $SU(N_f)_f$, and the baryon number conservation 
group $U(1)_B$ may be broken completely or to a certain
residual subgroup. In recent years there have been interesting works
studying the ground state of cold and dense quark matter, i.e., it 
has been investigated which color-superconducting phase is favored for
certain ranges of the quark chemical potential $\mu$. This question is 
also of phenomenological interest, since matter in the interior of 
neutron stars can reach densities up to an order of magnitude larger than the
nuclear ground state density while the temperature can be of the order of 
keV. Therefore, the core of a neutron star can be expected to be a color
superconductor and the question arises of which of the color-superconducting
phases it consists.  

For very large densities, where the quark masses of the $u$, $d$, and $s$
quarks can be considered degenerate since they are much smaller than the 
chemical potential, the ground state is the so-called
color-flavor-locked (CFL) phase \cite{alford}. In this phase, quarks of
all flavors and all colors form Cooper pairs, breaking spontaneously
$SU(3)_c\times SU(3)_f\times U(1)_{em}\times U(1)_B$ to a subgroup 
$SU(3)_{c+f}\times U(1)_{c+em}$. At smaller densities the 
situation is more complicated, because the strange mass cannot be neglected. 
Furthermore, the conditions of $\beta$-equilibrium and
electric charge neutrality impose restrictive conditions on the system. 
The simplest solution seems to be a pairing of only $u$ and $d$ quarks
\cite{bailin}, the so-called 2SC phase. In this phase, quarks of one
color remain unpaired, and the symmetry breaking pattern is  
$SU(3)_c\times SU(2)_f\times U(1)_{em}\times U(1)_B \to SU(2)_c\times SU(2)_f
\times U(1)_{c+em} \times U(1)_{em+B}$. In both CFL and 2SC phases the 
Fermi momenta of the quark flavors participating in pairing are assumed 
to be equal. This is a necessary condition for the conventional BCS
pairing mechanism, since in both cases quarks of different flavors 
form Cooper pairs. However, for moderate densities, this assumption is not 
valid. Therefore, the ground state is neither the (pure) CFL nor the 
(pure) 2SC state. 
Several other possibilities have been discussed. In principle, besides a 
transition to the normal-conducting state, there are
two classes of alternatives, both yielding color-superconducting states. 
In the first class, the difference in Fermi momenta is nonzero but small 
enough to still allow for pairing of quarks of different flavors. 
The second class accounts for cases in which quarks of the same flavor pair.

Let us first mention some options for the first class. First, there might be
a phase in which the Cooper pairs carry nonzero total momentum. In this 
case, the system exhibits a crystalline structure due to a spatially
varying energy gap \cite{alford2}. This kind of superconductivity is called 
LOFF phase and was first
discussed in solid-state physics \cite{loff}, where a difference in the 
electron Fermi momenta is induced by an external magnetic field.      
Second, there are studies about the ``gapless'' 2SC 
\cite{shovkovy} and CFL \cite{gCFL} phases. In these phases, at least a 
part of the 
quasiparticle excitations are ungapped, although the gap parameter 
is nonzero. This feature obviously can have enormous physical consequences,
for instance for the specific heat and the neutrino emissivity which 
both affect the cooling of a
neutron star with a core in a gapless color-superconducting phase. 
Other possible phases are derived from the CFL phase and contain kaon 
and/or eta condensates \cite{kaon}. Moreover, besides a displacement 
(LOFF phase), also a deformation of the Fermi surface has been discussed 
\cite{muether}.   

In this paper, we discuss the possibility of the second class, namely 
systems in which quarks of the same flavor form Cooper pairs
\cite{bailin,oneflavor,rischke,schaefer}. The simplest situation for this
kind of color superconductivity is a system of only 
one quark flavor. More realistic scenarios are two- or three-flavor 
systems where each quark flavor separately forms Cooper pairs. Another
possibility is a system where $u$ and $d$ quarks form a (gapless) 2SC
superconductor while the $s$ quarks pair separately. 

Since the attractive 
interaction of quarks is provided in the {\em anti}symmetric color antitriplet 
channel, the spin channel must be symmetric in order to ensure the
overall antisymmetry of the Cooper pair wave function. Consequently,
Cooper pairs consisting of quarks of the same flavor cannot carry total 
spin zero but must condense in the spin-one channel, $J=1$, where $J=L+S$
is the total spin of the Cooper pair, consisting of their angular momentum $L$
and their spin $S$. This ``two-triplet-condensation''
(color and spin triplets) renders the structure of the order parameter 
a complex $3 \times 3$ matrix. This is in contrast to the 2SC case, where, 
due to condensation in the color antitriplet, flavor and spin singlet 
channels, the order parameter is a complex 3-vector. In the CFL case, 
the order parameter is also a $3 \times 3$ matrix, originating from the color
and flavor antitriplets. 

Another system with this structure of the 
order parameter is superfluid $^3$He \cite{leggett,vollhardt}. 
In this nonrelativistic system, 
angular momentum $L$ and spin $S$ are separate quantum numbers, both giving
rise to a triplet structure of the condensate. In other
words, a Cooper pair of $^3$He atoms carries angular momentum one and spin
one. Without external magnetic fields, two phases of superfluid $^3$He 
are experimentally known, the A and B phases. The A phase is given by the 
order 
parameter structure $\Delta_{ij}=\d_{i3}(\d_{j1}+i\d_{j2})$, where the index
$i$ refers to spin and the index $j$ to angular momentum. This order
parameter spontaneously breaks $SO(3)_S\times SO(3)_L\times U(1)_N$ to
$U(1)_S\times U(1)_{L+N}$, where $SO(3)_S$ and $SO(3)_L$ are the spin and
angular momentum groups, respectively, and $U(1)_N$ is the particle number
conservation group. The B phase, 
which covers the largest region of the phase diagram, is given by
$\Delta_{ij}=\d_{ij}$, locking angular momentum with spin. Here, 
the residual group is $SO(3)_{S+L}$. While the gap function 
$\phi(\hat{\bf k})$ is isotropic in the B phase, it is anisotropic in the 
A phase. The condensation energy (density) $\Delta p$ of the superfluid states 
is given by the angular average of the square of the gap \cite{vollhardt}, 
\be \label{condenergy}
\Delta p = \frac{1}{2}\,N(0)\,\angk{\phi^2(\uk)}  \,\, ,
\ee
where $N(0)$ is the density of states at the Fermi surface and 
$\angk{-}\equiv\int\frac{d\Omega_k}{4\pi}$.
In weak coupling and at zero temperature, 
the ratio of the condensation energies of the A phase and
the B phase is (cf.\ Eqs.\ (3.72) and (3.75) of Ref.\ 
\cite{vollhardt})
\be \label{ratioAB}
\frac{\Delta p_A}{\Delta p_B} \simeq 0.88 \,\, .
\ee
Therefore, weak-coupling theory predicts the B phase to be the favored state.

It is the main goal of this paper to determine the favored state in a 
spin-one color superconductor. The paper is organized as follows:
In Sec.\ \ref{patterns}, we discuss possible symmetry breaking patterns
in a spin-one color superconductor. In general, the group 
$G=SU(3)_c\times SU(2)_J\times U(1)_{em} \times U(1)_B$ is spontaneously 
broken to a residual group $H$, where $SU(2)_J$ is the (relativistic)
spin group. A systematic list of order parameter matrices $\Delta$ 
and the corresponding residual groups is presented, based on the 
simple group-theoretical condition that $\Delta$ be invariant under 
transformations of $H$. Four phases with large residual groups are 
picked for further investigation in the next sections, namely the polar, 
planar, A, and CSL (color-spin-locked) phases. 

After establishing the formalism in Sec.\ \ref{definitions}, 
Sec.\ \ref{general1} is devoted to   
solving the gap equation for the general case of one or two nonzero (constant
or angular-dependent) energy gaps. The results are the gap functions for
zero temperature, $T=0$, at the Fermi surface 
\be
\phi_r(\uk)=\sqrt{\lambda_{k,r}}\,\phi_0 \,\, , 
\ee
where the angular dependence is contained in the quantities 
$\lambda_{k,r}\equiv \lambda_r(\hat{\bf k})$. The gap function occurs in the excitation 
spectrum of the quasiparticles,
\be
\e_{k,r} = \sqrt{(k-\mu)^2 + \lambda_{k,r}\,\phi_0^2} \,\, .
\ee
Let us briefly recall the situation 
of isotropic gap functions, $\lambda_{k,r}\equiv\lambda_r$. It has been shown that, in
this case and with $r = 1,2$, the gap parameter is given by 
\cite{rischke,QCDgapeq,son,ren,wang,schmitt}    
\be \label{gapconstant}
\phi_0 = 2\,\tilde{b}\,b_0'\,e^{-d} \,e^{-\zeta} \mu \, 
\exp\left(-\frac{\pi}{2\bar{g}}\right) \,\, ,
\ee
where  
\be \label{tildeb}
\bar{g}\equiv \frac{g}{3\sqrt{2}\pi}\,\, ,\qquad \tilde{b}\equiv 
256\pi^4\left(\frac{2}{N_f g^2}\right)^{5/2} \,\, , \qquad 
b_0'\equiv \exp\left(-\frac{\pi^2 + 4}{8}\right) \,\, ,
\ee
with the strong coupling constant $g$. The exponent $d$ is zero in all
spin-zero phases and nonzero for $J=1$. The exponent $\zeta$ is defined 
as
\be \label{zetaiso}
\zeta\equiv \ln (\lambda_1^{a_1}\lambda_2^{a_2})^{1/2} \,\, ,
\ee
where the numbers $a_1$ and $a_2$ have to 
be determined for each phase separately. They fulfil the condition 
\be \label{condition}
a_1 + a_2 = 1 \,\, .
\ee
The notation $e^{-\zeta}$ instead of $(\lambda_1^{a_1}\lambda_2^{a_2})^{-1/2}$ will
turn out to be convenient for the generalization to the case of 
anisotropic gap functions.

In the case of the 2SC phase, there is only one gapped quasiparticle 
excitation branch, $a_1=\lambda_1=1$, $a_2=\lambda_2=0$, and hence
\be
\phi_0^{\rm 2SC} = 2\,\tilde{b}\,b_0'\,\mu \, 
\exp\left(-\frac{\pi}{2\bar{g}}\right)
\,\, ,
\ee
whereas in the CFL phase, there are two different gaps, $a_1=1/3$, $a_2=2/3$, 
$\lambda_1=4$, $\lambda_2=1$, which leads to $\phi_0^{\rm CFL} = 
2^{-1/3} \,\phi_0^{\rm 2SC}$.

In Sec.\ \ref{general2}, we compute the transition temperature $T_c$
for the transition from the normal-conducting to the superconducting state.
As for the gap, this is a generalization of the cases with constant 
(= angular-independent) gaps, for which the transition temperature
is given by \cite{schmitt}
\be \label{tempconstant}
\frac{T_c}{\phi_0} = \frac{e^\g}{\pi} e^\zeta 
\simeq 0.57 \, e^{\zeta} \,\, ,
\ee
where $\g\simeq 0.577$ is the Euler-Mascheroni constant. It has been 
one of the main conclusions of Ref.\ \cite{schmitt} that this expression shows
the violation of the well-known BCS relation $T_c/\phi_0 \simeq 0.57$
in the case of a two-gap structure, i.e., $\lambda_{1,2}$, $a_{1,2}\neq 0$. 
(Also in the case of the gapless 2SC phase, this relation is violated
\cite{shovkovy}.) 
 
As in the case of $^3$He, one expects the preferred phase to have the 
largest condensation energy, cf.\ Eq.\ (\ref{condenergy}). We specify this 
statement in Sec.\ \ref{general3} with a general derivation of the 
pressure in an arbitrary color-superconducting phase. 

In Sec.\ \ref{results}, we determine the excitation spectrum, the gap 
functions, the critical temperature, and the pressure for the 
polar, planar, A, and CSL phases. For each phase we consider three special 
cases, termed ``longitudinal'', ``mixed'', and ``transverse''. These three
cases arise from the following property of spin-one phases: Contrary to 
a spin-0 color superconductor, where only quarks of the same chirality
form Cooper pairs (RR and LL pairs), in a spin-1 color superconductor
also pairing of quarks with opposite chirality (RL and LR pairs) is possible
\cite{rischke,schaefer,schmitt}. In general, the order parameter contains
a linear combination of both kinds of condensates. We focus on the cases
of pure RR/LL condensates (longitudinal), a special admixture of RR/LL
and RL/LR condensates (mixed), and pure RL/LR condensates (transverse).
Consequently, in total 12 phases are studied. 

In Sec.\ \ref{conclusions} we summarize the results and give an outlook 
for possible consequences in neutron stars with color-superconducting
cores.

Our convention for the metric tensor is 
$g^{\mu\nu}=\mbox{diag}\{1,-1,-1,-1\}$. 
Our units are $\hbar=c=k_B=1$. Four-vectors
are denoted by capital letters, 
$K\equiv K^\mu=(k_0,{\bf k})$, 
and $k\equiv|{\bf k}|$, while $\uk\equiv{\bf k}/k$.
We work in the imaginary-time formalism, i.e., $T/V \sum_K \equiv
T \sum_n \int d^3{\bf k}/(2\pi)^3$, where $n$ labels the Matsubara 
frequencies $\omega_n \equiv i k_0$. For bosons, $\omega_n=2n \pi T$,
for fermions, $\omega_n=(2n+1) \pi T$.

\section{Patterns of symmetry breaking} \label{patterns}

\renewcommand{\labelenumii}{(\roman{enumii})}

In this section, we discuss possible symmetry breaking patterns in a spin-one
color superconductor. In other words, we present a systematic classification 
of theoretically possible superconducting phases. In the case
of a one-flavor, spin-one color superconductor, the relevant original 
symmetry group of the system is
\be
G=G_1\times G_2\times G_3 \,\, ,
\ee
where 
\be
G_1 = SU(3)_c \,\, , \qquad G_2 = SU(2)_J \,\, ,\qquad G_3 = U(1)_{em} \,\, .
\ee
Note that the global group $U(1)_B$, accounting for baryon number conservation,
has the same generator as the local symmetry group $U(1)_{em}$. This is not 
true in a system with $N_f>1$, when at least two quark flavors differ in their 
electric charge. In particular, in the CFL phase ($N_f=3$), this leads to
the fact that the system is not an electromagnetic superconductor but a 
superfluid (the breakdown of $U(1)_B$ gives rise to a Goldstone boson). 
For $N_f=1$, however, these two phenomena are coupled, i.e., a 
superflow is equivalent to a supercurrent. 
 
The order parameter $\Delta$ is
an element of a representation of $G$. In the following, we use the term
order parameter somewhat sloppily for the pure matrix structure $\Delta$. 
For a spin-one color superconductor, the relevant representation of
$G$ is the 
tensor product of the antisymmetric color antitriplet $[\bar{\bf 3}]_c^a$ 
and the symmetric spin triplet $[{\bf 3}]_J^s$, 
\be \label{representation}
\D \in  [\bar{\bf 3}]_c^a \otimes [{\bf 3}]_J^s \,\, .
\ee
Therefore, $\D$ is, as in the case of superfluid $^3$He, a complex $3\times 3$
matrix. There is no nontrivial contribution from the flavor 
structure since we consider systems with only one quark flavor.
The group $G$ is spontaneously broken down to a residual (proper) subgroup 
$H\subseteq G$.
This means that any transformation $g\in H$ leaves the order parameter 
invariant,
\be \label{invariance0}
g(\D) = \D \,\, .
\ee
In the following, we investigate this invariance condition in order to 
determine all possible order parameters $\D$ and the corresponding 
residual groups $H$. The method we use in this section is motivated 
by the analogous one for the case of superfluid $^3$He \cite{vollhardt}.

First, one has to specify how $G$ acts on the order 
parameter in Eq.\ (\ref{invariance0}). To this end, we write an arbitrary 
group element $g\in G$ as (in this section, no confusion with the 
strong coupling constant $g$ is possible)
\be
g=(g_1,g_2,g_3) \,\, ,
\ee
where
\be \label{g1g2g3}
g_1=\exp(-ia_mT_m^T) \,\, , \qquad g_2=\exp(ib_nJ_n) \,\, , \qquad 
g_3=\exp(2ic\bf{1}) \,\, , 
\ee
with real coefficients $a_m$ ($m=1,\ldots ,8$), $b_n$ ($n=1,2,3$), and $c$. 
The Gell-Mann matrices $T_m$ generate the group $SU(3)_c$, and we 
have taken into account that the color representation is 
an {\it anti}triplet.
The matrices $J_n$ are the generators of the spin group 
$SU(2)_J$, $(J_n)_{ij}=-i\,\e_{nij}$. For the generator of $U(1)$ we 
choose $2\cdot{\bf 1}$, where
${\bf 1}$ is the $3\times 3$ unit matrix. The factor 2 accounts for the
diquark nature of the order parameter. 

Let us now introduce a basis $J_i\otimes \kappa_j$ ($i,j=1,2,3$) for the 
representation given in Eq.\ (\ref{representation}). For the basis elements
of $[\bar{\bf 3}]_c^a$ we choose the antisymmetric $3\times 3$ matrices
$J_i$, as introduced above as generators of the spin group. The basis
of $[{\bf 3}]_J^s$ is given by the 3-vector $\kappa_j$, which will be
specified in Sec.\ \ref{results}, cf.\ Eq.\ (\ref{Mk}). Thus, we have 
to consider the action of $G$ on 
\be \label{Mgeneral}
{\cal M}_{\bf k} \equiv J_i\,\Delta_{ij}\kappa_j \,\, .
\ee
We have 
\be
g({\cal M}_{\bf k}) = g_3 \, g_1^{ik}\, J_k \,\D_{ij} \,g_2^{j\ell} \,\kappa_\ell 
\,\, .
\ee
Therefore, the matrix $\D$ transforms as
\be
g(\D_{ij}) = g_3 \, g_1^{ki} \, \D_{k\ell} \, g_2^{\ell j} \,\, .
\ee
Then, using Eqs.\ (\ref{g1g2g3}), the infinitesimal transformations of 
$\D$ by $G$ are given by
\be
g(\D) \simeq \D  - a_mT_m\D + b_n\D J_n + 2c\,\D\,\, ,
\ee
where $T_m\D$ as well as $\D J_n$ are matrix products. 
The invariance condition for the order parameter
(\ref{invariance0}) is thus equivalent to
\be  \label{invariance}
- a_mT_m\D + b_n\D J_n + 2c\,\D = 0\,\, .
\ee
This matrix equation can be written as a system of nine equations for
the nine complex entries $\D_{11},\ldots,\D_{33}$ of the matrix $\D$. 
In principle,
one can find all possible symmetry breaking patterns and corresponding order
parameters by setting the determinant of the coefficient matrix to zero. Then,
each possibility to render the determinant zero yields a set of 
conditions for the coefficients $a_m$, $b_n$, $c$, and it can be checked if 
these conditions correspond to a residual subgroub $H$. But since this is 
much too complicated, we proceed
via investigating possible subgroups explicitly. In the following, we 
focus only on the continuous subgroups of $G$. 

Let us start with subgroups 
$H$ that contain the smallest possible continuous group, $U(1)$, i.e., 
\be \label{H'}
H=U(1)\times H' \,\, ,
\ee
where $H'$ is a direct product of Lie groups.
The residual $U(1)$ must be generated by a $3\times 3$ matrix $U$ which is 
a linear 
combination of the generators of $G$, i.e., in general,
\be
U=a_mT_m + b_nJ_n + 2c{\bf 1} \,\, .
\ee
Let us restrict ourselves to linear combinations that involve one generator of each 
group $G_1$, $G_2$, $G_3$, for instance  
\be  \label{generator1}
U = a_8T_8 + b_3J_3 + 2c{\bf 1} \,\, .
\ee 
With Eq.\ (\ref{generator1}), the invariance condition
\be \label{invarianceapp}
e^{iU}(\D)=\D
\ee
results in a system of nine equations, which can be discussed explicitly. This
is done in Appendix \ref{orderparameters}. We find ten different 
order parameters, eight of them depending on two or more parameters 
$\Delta_1, \Delta_2, \ldots$, which physically corresponds to 
more than one gap function. With the normalization 
\be \label{normalize2}
\Tr(\D\D^\dag)=1 \,\, ,
\ee
the calculation in Appendix \ref{orderparameters} yields also two matrices
$\Delta$ that do not depend on any free parameter (which corresponds to 
only one gap function). In analogy to $^3$He, let us call 
the phases defined by this kind of order parameter {\em inert}
\cite{vollhardt}. All experimentally known states of superfluid 
$^3$He belong to this class of order parameters. 
Mathematically speaking, these matrices play a special role due to a theorem 
(``Michel's Theorem'') \cite{vollhardt,michel}, which ensures that these 
order parameters correspond to a stationary point of any $G$-invariant 
functional of $\D$ (for instance the effective potential). 
This is the reason why we also focus on these order parameters in the 
explicit calculations of the physical properties, see Sec.\ \ref{results}.

\begin{figure}[ht] 
\begin{center}
\includegraphics[width=11cm]{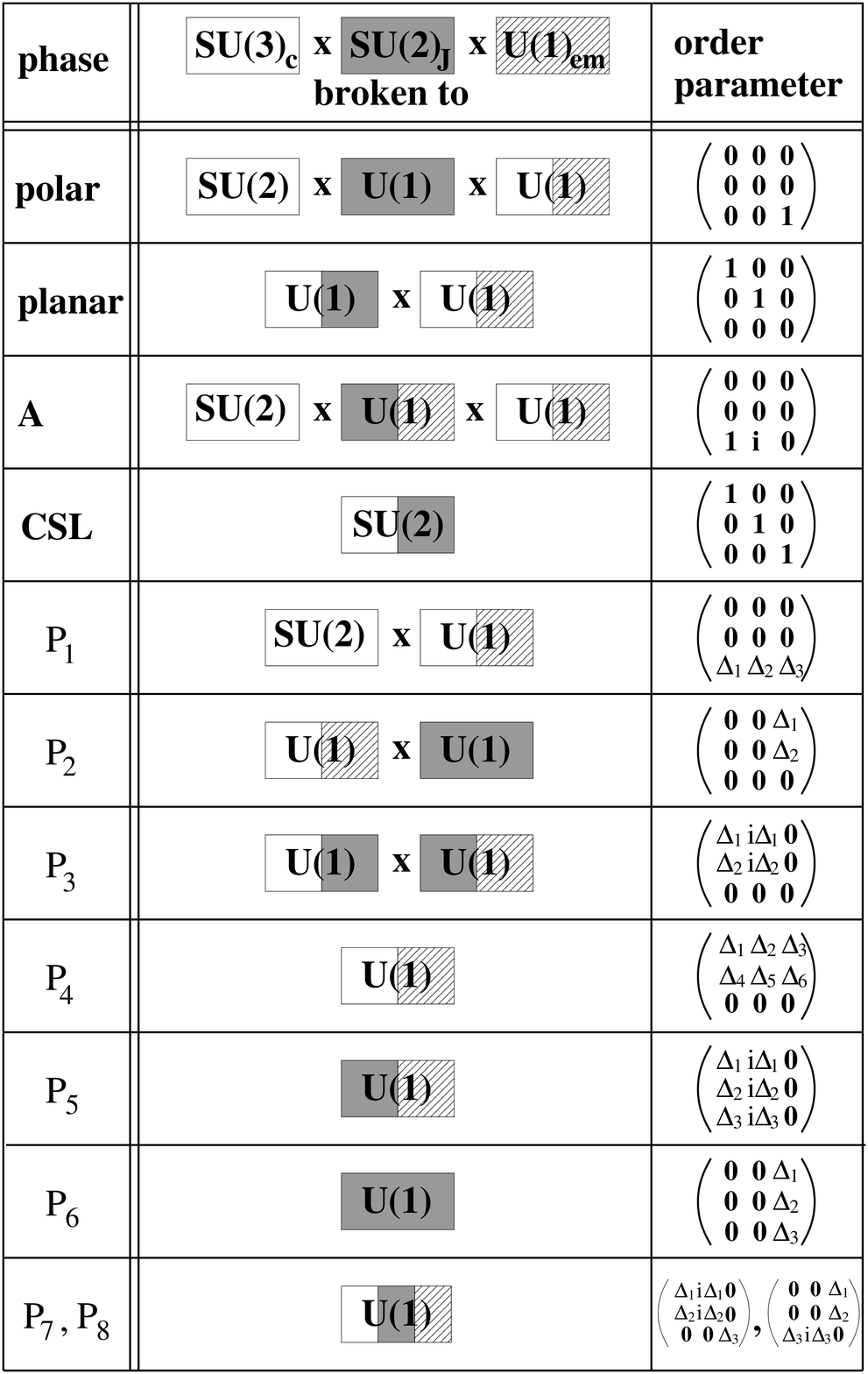}
\vspace{0.5cm}
\caption[Symmetry breaking patterns in a spin-one color superconductor]
{Symmetry breaking patterns and 
order parameters for a 
spin-one color superconductor. The original symmetry group (first line) 
is given 
by the color gauge group $SU(3)_c$ (blank background), the spin group 
$SU(2)_J$ (grey background), and the electromagnetic gauge group $U(1)_{em}$ 
(hatched background). The backgrounds of the 
residual groups illustrate the symmetry breaking pattern. For
instance, the blank $SU(2)$ occuring in the residual group of the
polar phase is generated solely by generators of the original $SU(3)_c$ 
while the blank/hatched $U(1)$ in the same line is generated by a 
linear combination of the generators of the original $SU(2)_J$ and $U(1)_{em}$ 
groups. For the explicit expression of these generators, see text.
}
\label{symmetries}
\end{center}
\end{figure}

One order parameter, found with the ansatz (\ref{generator1}) and 
corresponding to an inert phase, is  
\be
\D=\left(\begin{array}{ccc} 0&0&0\\0&0&0\\0&0&1 
\end{array}\right) \,\, , \label{orderpolar1} 
\ee
defining the {\em polar phase}. It has its analogue in 
$^3$He \cite{vollhardt}, where the order parameter matrix is identical.
For the case of a color superconductor, certain aspects of the polar phase 
have already been discussed in Refs.\  
\cite{schaefer,schmitt,schmitt2,schmitt3}.
For the corresponding residual group $H$ see Fig.\ \ref{symmetries}, where
we list all order parameters. More details, especially the explicit forms
of the generators of $H$, are given in Appendix \ref{orderparameters}.
The second order parameter giving rise to an inert phase is
\be
\D= \frac{1}{\sqrt{2}}\left(\begin{array}{ccc} 0&0&0\\ 0&0&0\\ 1&i&0 
\end{array}\right) \,\, , \label{orderA1} 
\ee
This order parameter leads to the {\em A phase} \cite{vollhardt,schaefer}.

In order to find all possible (inert) order parameters, it is necessary to 
consider at least one more combination for the residual $U(1)$ in 
Eq.\ (\ref{H'}), namely 
\be  \label{generator2}
U = a_2T_2 + b_3J_3 + 2c{\bf 1} \,\, .
\ee 
The reason why $T_2$ plays a special role is that we 
used the generator $J_3$ of the spin group, which is proportional to
$T_2$. Consequently, we expect to find additional residual groups that 
connect the color group with the spin group
(meaning a residual $U(1)$ generated by a combination of a color and a 
spin generator). The calculations with this generator are completely 
analogous to the ones with the ansatz (\ref{generator1}). Therefore, we 
present the result without elaborating on the details.   
The ansatz (\ref{generator2}) yields one inert order parameter that is 
different from the above ones, namely
\be \label{orderplanar}
\D=\frac{1}{\sqrt{2}}\left(\begin{array}{ccc} 1&0&0\\0&1&0\\0&0&0 
\end{array}\right) \,\, .
\ee
This order parameter corresponds to the {\em planar phase} 
\cite{vollhardt,schaefer}. The residual group is $H=U(1)\times U(1)$,
with generators
\be
U=2T_2 + J_3 \,\, ,\qquad V=T_8+\frac{1}{4\sqrt{3}} \,{\bf 1} \,\, .
\ee 

Let us now turn to possible groups $H$ that do not contain any $U(1)$ but
solely consist of higher-dimensional Lie groups, say 
\be \label{H'2}
H=SU(2)\times H' \,\, .
\ee
Let $U_1$, $U_2$, $U_3$ be the generators of the residual $SU(2)$. They 
are linear combinations of the generators of $G$,
\be
U_i=a^i_mT_m + b^i_nJ_n + 2c^i{\bf 1} \,\, , \qquad i = 1,2,3 \,\, .
\ee
Since they must fulfil
the $SU(2)$ commutation relations,  
\be \label{SU2commutation}
[J_i,J_j]=i\,\e_{ijk} J_k \,\, , \qquad i,j,k \le 3 \,\, .
\ee  
they must not contain the generator of 
$G_3=U(1)$, the unit matrix, i.e., $c^1=c^2=c^3=0$ . Therefore, there are 
three possibilities.
First, each $U_i$ is a combination of color and spin 
generators. Second and third, each $U_i$ is composed solely of color or 
spin generators, respectively. The simplest options to realize these 
cases are
\begin{subequations} \label{SU2gen}
\bea 
&& U_i=T_i'+J_i \,\, , \label{casea}\\
&& U_i=T_i' \,\, ,    \label{caseb}\\
&& U_i=J_i \,\, ,  \label{casec} 
\eea
\end{subequations}
where $(T_1',T_2',T_3')$ is either given by $(T_1,T_2,T_3)$ or  
$(2T_7,-2T_5,2T_2)$, which both fulfil the required commutation relations. 
Using the options (\ref{casea}) -- (\ref{casec}), let us first show that 
$H'$ in Eq.\ (\ref{H'2}) cannot be a second $SU(2)$. To this end, assume
that $H'=SU(2)$ with generators $V_1$, $V_2$, $V_3$, which have the same form
as the generators $U_i$ in Eqs.\ (\ref{SU2gen}). Then, since
the Lie algebra of $H$ is a direct sum of the constituent Lie algebras,
we have to require
\be
[U_i,V_j] = 0\,\, .
\ee
This condition reduces all options to one, namely
\be
U_i = T_i' \,\, , \qquad V_i = J_i 
\ee
(or vice versa). However, now the invariance equation for the order parameter yields
\be \label{exclude}
\D J_i=0 \,\, ,
\ee
for all $i=1,2,3$, which does not allow for a nonzero order parameter $\D$.
Therefore, $H'=SU(2)$ is forbidden. Since the cases with $H'=U(1)$ and
$H'=U(1)\times U(1)$ were already covered in the above discussion,
the only possibility that is left is $H'={\bf 1}$ and thus $H=SU(2)$. 
 
The generators in Eq.\ (\ref{casec}) can immediately be excluded since they 
also lead to Eq.\ (\ref{exclude}). The same 
argument excludes case (\ref{caseb}) with $(T_1',T_2',T_3')=(2T_7,-2T_5,2T_2)$.
Case (\ref{caseb}) with $(T_1',T_2',T_3')=(T_1,T_2,T_3)$ leads to two 
order parameters already considered above, namely the polar phase, 
Eq.\ (\ref{orderpolar1}), and the A phase, Eq.\ (\ref{orderA1}). 
In case (\ref{casea}), only $(T_1',T_2',T_3')=(2T_7,-2T_5,2T_2)$ is possible. 
With
\be
U_i\D = -T_i'\D + \D J_i = 0 \,\, ,
\ee
one finds
\be \label{orderCSL}
\D=\frac{1}{\sqrt{3}}\left(\begin{array}{ccc} 1&0&0\\0&1&0\\0&0&1 
\end{array}\right) \,\, . \\      
\ee 
Indeed, it can be checked with Eq.\ (\ref{invariance}) that this order
parameter leads to 
\be
a_1=a_3=a_4=a_6=a_8=c=0\,\, , \quad a_2 = 2b_3  \,\, , \quad a_5 = -2b_2
\,\, , \quad a_7 = 2b_1 \,\, ,
\ee
which corresponds to $H=SU(2)$, consisting of joint rotations in color and 
spin space. This is the {\em color-spin-locked (CSL) phase}, discussed for
a spin-one color superconductor in Refs.\ 
\cite{bailin,schaefer,schmitt,schmitt2,schmitt3}. 
It is the analogue of the B phase in superfluid $^3$He.

Finally, we give an argument why an even larger subgroup, i.e., an 
$SU(3)$, cannot occur in the residual group $H$. Assume that there 
are eight generators $W_1,\ldots ,W_8$ of this $SU(3)$. Then, as 
for the $SU(2)$ subgroup above, there can be no contribution to 
$W_1,\ldots ,W_8$ from the 
$G_3$ generator due to the $SU(3)$ commutation relations for the generators,
\be \label{SU3commutation}
[W_i,W_j] = i\,f_{ijk} W_k \,\, , \qquad i,j,k \le 8 \,\, ,
\ee
where $f_{ijk}$ are the $SU(3)$ structure constants.
Also, $W_i=T_i$ is excluded 
because in this case the invariance condition yields 
$T_8\D=0$ and thus $\D=0$. Therefore, at least one of the spin 
generators has to be included. For instance, choose $W_8=T_8 + b_3 J_3$.
Then, from the commutation relation \mbox{$[W_4,W_5]\sim W_8$} we conclude
that also $J_1$ and $J_2$ must be included via $W_4=T_4 + b_1 J_1$ and   
$W_5=T_5 + b_2 J_2$. But now the three equations $W_4\D=W_5\D=W_8\D=0$
lead to $\D=0$. Therefore, we conclude that
there is no residual group $H$ that contains an $SU(3)$. For a more 
rigorous proof one has to take into account more complicated linear 
combinations of the original generators. 

In Fig.\ \ref{symmetries}, we summarize our results in a list 
of all superconducting phases that we have found in the above 
discussion. It should be mentioned that this list is not complete, since
for the generators of the residual $U(1)$'s we have restricted ourselves to 
two special forms given in Eqs.\ (\ref{generator1}) and (\ref{generator2}).
Therefore, there are certainly more (at least non-inert) order parameters 
that lead to an allowed symmetry breaking. The 
inert states are listed in the first four lines. Each of these four states 
has its analogue in superfluid $^3$He. Note that the
${\rm A}_1$ phase, which is experimentally observed in $^3$He in the 
presence of an external magnetic field, does not lead to an allowed 
symmetry breaking
in the case of a spin-one color superconductor \cite{schaefer}. 
To see this, one inserts the order parameter of the ${\rm A}_1$ phase,
\be
\D=\frac{1}{2}
\left(\begin{array}{ccc} 1 & i & 0 \\ -i & 1 & 0 \\
0 & 0 & 0 \end{array}\right) \,\, ,
\ee
into Eq.\ (\ref{invariance}). One obtains
\be
a_1=a_3=b_1=b_2=0\,\, ,\quad a_4=a_7\,\, ,\quad a_5=a_6\,\, , \quad
\frac{1}{2}a_2+\frac{1}{2\sqrt{3}}a_8-b_3-2c=0 \,\, .
\ee
These 7 conditions suggest that ${\rm dim}\,H=12-7=5$ and thus 
$H=SU(2)\times U(1)\times U(1)$. However, there is no possibility to construct
three generators from the above conditions that fulfil the $SU(2)$ 
commutation relations. For instance, assume that two of these 
generators
are given by $U_1=T_4 + T_7$ and $U_2=T_5 +T_6$. Then, with 
Eq.\ (\ref{SU3commutation}), $[U_1,U_2]\sim T_3$. But since $a_3=0$, 
the third generator $U_3$ cannot be proportional to $T_3$. 
Consequently, there is no ${\rm A}_1$ phase in a spin-one color superconductor.
  
Below the four inert states we list the eight non-inert states which 
have been found in Appendix \ref{orderparameters} and which we term 
${\rm P}_1, \ldots , {\rm P}_8$. Note that one of these 
non-inert phases, ${\rm P}_1$, has a larger residual symmetry group than the 
planar and CSL phases. 

There are several properties of the spin-one phases which can easily
be read off from Fig.\ \ref{symmetries}. First, consider  
the spin group $SU(2)_J$. This symmetry accounts for the rotational symmetry
in real space of the normal-conducting phase (in the case of 
the spin-one representation, one can equivalently consider $SO(3)_J$ instead 
of $SU(2)_J$). The list shows that spatial symmetry is broken in each case. 
For instance, in the polar phase, $SU(2)_J$ is broken to its subgroup $U(1)_J$.
Therefore, the superconducting phase is invariant under rotations around
one fixed axis in real space. In most
of the other cases, the breaking of the spatial rotation symmetry is more
subtle: For instance in the planar phase, the superconducting state is 
invariant under a special joint rotation in color and real space. 
The most interesting breakdown of spatial symmetries is present in the CSL phase. 
Here, any rotation in real space leaves the system invariant as long as one 
simultaneously performes the same rotation in fundamental color space 
which is spanned by the three directions red, green, and blue. 

Next, let us read off some properties concerning the color symmetry.
It is obvious that in none of the cases the 
full color symmetry is preserved. In this sense, it is justified to call
each phase a color superconductor. 
In three of the cases, there is a residual 
color subgroup $SU(2)$, namely in the polar phase, the A phase, and the
${\rm P}_1$ phase. Mathematically speaking, 
this residual group originates from
the fact that the order parameter has only nonzero elements in its third
row. Therefore, the third direction in fundamental color space is preferred.
Physically, this means that the Cooper pairs carry color charge anti-blue, 
or, in other words, only red and green quarks form Cooper pairs. Of course,
the choice of the anti-blue direction is convention; more generally speaking,
quarks of one color remain unpaired. Remember that this is also true 
for the 2SC phase. 

A spontaneously broken gauge symmetry gives rise to massive gauge bosons. 
In the case
of a color superconductor, these masses are the magnetic screening masses 
of the gluons. Therefore, in the cases where there is a residual color
subgroup $SU(2)$, we expect a Meissner effect for five of the eight 
gluons. Three of the gluons, however, namely those corresponding to 
the generators $T_1$, $T_2$, $T_3$, do not attain a Meissner mass. This is 
also obvious from the fact 
that these gluons do not see the (anti-)blue color charge which is
carried by the Cooper pairs. 
Also with respect to the breakdown of the color symmetry, the CSL 
phase is exceptional. Although there is a residual $SU(2)$, where three
of the color generators are involved, we expect all eight gluons to 
attain a Meissner mass. To this end, note that this residual $SU(2)$ is 
a global symmetry and 
therefore all dimensions of the gauge group have to be considered as broken.   
This is analogous to the CFL phase, which also exhibits a color Meissner 
effect for all eight gluons \cite{meissner3}. For a more detailed and 
quantitative discussion of the color Meissner effect, 
see Ref.\ \cite{schmitt3}.

In order to discuss the question whether the color superconductors in Fig.\ 
\ref{symmetries} are also electromagnetic superconductors, one has to consider
the hatched backgrounds in the residual groups. From ordinary 
superconductors, we know
that $U(1)_{em}$ is spontaneously broken below the critical temperature.
Therefore, a simple conclusion is that all states in the list without a 
hatched background occurring in the residual group are electromagnetic
superconductors. Obviously, this is the case for the CSL phase and for one of 
the non-inert states, namely the ${\rm P}_6$ phase. The electromagnetic
Meissner effect in the CSL and polar phases has been discussed in detail 
in Refs.\ \cite{schmitt2,schmitt3}. In particular, it has been shown that
also the polar phase exhibits an electromagnetic Meissner effect in 
the case of a many-flavor system, where at least two of the 
quark flavors have nonequal electric charges.

\section{Gaps, transition temperature, and pressure} \label{effpot}

In this section, we derive general expressions for the gaps 
(Sec.\ \ref{general1}), the
transition temperature (Sec.\ \ref{general2}), and the pressure 
(Sec.\ \ref{general3}), valid for all 
color-superconducting phases we consider. In Sec.\ \ref{results} we
evaluate these expressions for several spin-one color superconductors.

\subsection{Notations and definitions} \label{definitions}

The starting point is the effective action $\G$, which yields the gap equation
as well as the effective potential $V_{\rm eff}\equiv -\frac{T}{V}\G$ and the 
pressure $p$, which is the negative of the effective potential at its 
stationary point, $p=-V_{\rm eff}$.

The effective action $\Gamma$ can be derived from the QCD partition function 
using the Cornwall-Jackiw-Tomboulis (CJT) formalism \cite{cjt}. 
The resulting functional can be written as \cite{review,abuki,ruester}
\be \label{Gamma1}
\G[D_G,D_F] = -\frac{1}{2}\Tr\ln D_G^{-1}
                        -\frac{1}{2}\Tr\left(\Delta_0^{-1}D_G-1\right)
                        +\frac{1}{2}\Tr\ln D_F^{-1}
                        +\frac{1}{2}\Tr\left({\cal S}_0^{-1}D_F-1\right)
                        +\G_2[D_G,D_F] \,\, .
\ee
Here, $D_G$ ($\Delta_0$) and $D_F$ (${\cal S}_0$) are the full (tree-level) 
gluon and fermion propagators, respectively. All propagators
are defined in Nambu-Gor'kov space. We account for the doubling of degrees
of freedom in this basis by introducing the factor $1/2$. 
The traces run over Nambu-Gor'kov, Dirac, flavor, 
color, and momentum space, and $\G_2[D_G,D_F]$ denotes the sum of all
two-particle irreducible diagrams. In the following, we will consider
the two-loop approximation of this sum which, for the fermionic degrees of
freedom, is equivalent to taking into account only one diagram 
(the ``sunset'' diagram \cite{ruester}).
The stationarity conditions for $\G[D_G,D_F]$ yield 
Dyson-Schwinger equations for the inverse gluon and fermion propagators,
\begin{subequations}
\bea \label{DS}
\Delta^{-1} &=& \Delta_0^{-1} + \Pi \,\, , \label{DS1} \\
{\cal S}^{-1} &=& {\cal S}_0^{-1} + \Sigma \,\, . \label{DS2}
\eea
\end{subequations}
Here, the pair of propagators $(\Delta,{\cal S})$ is the stationary point of 
the effective potential and we defined the gluon and fermion self-energies as
\be
\Pi\equiv-2\left.\frac{\d\G_2}{\d D_G}\right|_{(D_G,D_F)=(\Delta,{\cal S})}
\quad, \qquad \Sigma\equiv\left.2\frac{\d\G_2}{\d D_F}
\right|_{(D_G,D_F)=(\Delta,{\cal S})} \,\, .
\ee 
The free inverse fermion propagator is 
\be \label{free}
{\cal S}_0^{-1}=\left(\begin{array}{cc} \left[G_0^+\right]^{-1} & 0 \\ 
                                 0 &  \left[G_0^-\right]^{-1}
               \end{array}\right) \,\, ,
\ee
where $[G_0^+]^{-1}$ and $[G_0^-]^{-1}$ are the free inverse propagators
for massless quarks and massless charge-conjugate quarks in the presence of 
a chemical potential $\m$,
\be 
[G_0^\pm]^{-1} = \g^\m K_\m \pm \m \g_0 \,\, .
\ee
In order to find the full propagators, one has to solve the Dyson-Schwinger
equations, Eqs.\ (\ref{DS}), self-consistently. To this end, 
we denote the entries of the fermion self-energy in Nambu-Gor'kov space by
\be \label{sigmanambu} 
\Sigma\equiv\left(\begin{array}{cc}\Sigma^+ & \Phi^- \\ \Phi^+ & \Sigma^- 
\end{array}\right) \,\, ,
\ee
and invert Eq.\ (\ref{DS2}) formally 
\cite{manuel}, which yields the full quark propagator in the form
\be \label{fullquark}
{\cal S} = \left(\begin{array}{cc} G^+ & \Xi^- \\ \Xi^+ & G^- 
\end{array}\right) \,\, ,
\ee   
where the fermion propagators for quasiparticles and charge-conjugate 
quasiparticles are
\be \label{Gpm}
G^\pm = \left\{[G_0^\pm]^{-1} + \Sigma^\pm - 
\Phi^\mp([G_0^\mp]^{-1}+\Sigma^\mp)^{-1}
\Phi^\pm\right\}^{-1} \,\, ,
\ee 
and the so-called anomalous propagators, typically nonzero for a 
superconducting system, are given by
\be \label{Xpm}
\Xi^\pm = -([G_0^\mp]^{-1} + 
\Sigma^\mp)^{-1}\Phi^\pm G^\pm \,\, .
\ee
In the two-loop approximation of $\G_2$, the quark self-energy $\Sigma(K)$
in momentum space is 
\be \label{sigmaeq}
\Sigma(K) = -g^2 \frac{T}{V}\sum_Q \G^\m_a \, {\cal S}(Q) \, \G_b^\n \, 
\D_{\m\n}^{ab}(K-Q) \,\, ,
\ee
where 
\be \label{Gammadef}
\G_a^\m \equiv \left(\begin{array}{cc} \g^\m T_a & 0 \\ 0 & -\g^\m 
T_a^T\end{array}
\right) \,\, ,
\ee
with the Gell-Mann matrices $T_a$, $a=1,\ldots, 8$.
Due to the Nambu-Gor'kov structure, Eq.\ (\ref{sigmaeq}) is actually a set of 
four 
equations. With Eqs.\ (\ref{sigmanambu}) and (\ref{fullquark}), the 
off-diagonal (21)-component leads to the gap equation 
\be
\Phi^+(K) = g^2\frac{T}{V}\sum_Q \g^\m \, T_a^T \, \Xi^+(Q) 
\, \g^\n \, T_b \, \D_{\m\n}^{ab}(K-Q) \label{gapeq21} \,\, .
\ee
The quantities $\Phi^\pm(K)$ are matrices in flavor, color, 
and Dirac space and functions of the quark four-momentum $K$. 
Both quantities are related via 
\be \label{phiminus}
\Phi^- = \g_0 [\Phi^+]^\dag\g_0 \,\, .
\ee
Following Ref.\ \cite{schmitt}, we term the matrix $\Phi^+(K)$ {\em gap matrix} 
and use, for condensation in the even-parity channel
and in the ultrarelativistic limit, the ansatz
\be \label{gapmatrix}
\Phi^+(K)=\sum_{e=\pm}\phi_e(K)\,{\cal M}_{\bf k}\,\Lambda_{\bf k}^e \,\, . 
\ee
The Dirac matrices $\Lambda_{\bf k}^e\equiv(1+e\g_0\vg\cdot\uk)/2$, where $e=\pm$, 
are projectors onto positive and negative energy states, and $\phi_e(K)$ 
is the gap function. The quantity 
${\cal M}_{\bf k}$ is a matrix in color, flavor, and Dirac space. 
It is defined by the order 
parameter and thus determines the color-superconducting phase. For the explicit
form of ${\cal M}_{\bf k}$ in the case of a spin-one color superconductor, see
Eq.\ (\ref{Mk}).     
We can always choose ${\cal M}_{\bf k}$ such
that it commutes with the energy projectors, 
\be \label{M}
[{\cal M}_{\bf k},\Lambda_{\bf k}^e]=0 \,\, .
\ee   
The diagonal elements of the quark self-energy 
can be approximated as \cite{ren,wang}
\be \label{sigmaapp}
\Sigma^+=\Sigma^-\simeq \g_0 \bar{g}^2k_0\,\ln\frac{M^2}{k_0^2} \,\, ,
\ee
where $M^2=(3\pi/4)m_g^2$; the zero-temperature
gluon mass parameter (squared) is $m_g^2=N_fg^2\m^2/(6\pi^2)$.

Using Eqs.\ (\ref{Gpm}), (\ref{gapmatrix}), and (\ref{sigmaapp}),   
we find for the fermion propagators
\be \label{prop}
G^\pm = \left(\left[G_0^\mp\right]^{-1} + \Sigma^\mp\right)\sum_{e,r}
{\cal P}_{{\bf k},r}^\pm\, \Lambda_{\bf k}^{\mp e} 
\frac{1}{\left[k_0/Z(k_0)\right]^2 - (\e_{{\bf k},r}^e)^2} \,\, ,
\ee
where 
\be 
Z(k_0) \equiv \left(1 + \bar{g}^2\ln\frac{M^2}{k_0^2}\right)^{-1}
\ee
is the wave function renormalization factor introduced in Ref.\ 
\cite{manuel}. 
In Eq.\ (\ref{prop}) we introduced two sets of projectors, 
${\cal P}_{{\bf k},r}^+$ and ${\cal P}_{{\bf k},r}^-$; 
they project onto the eigenspaces of the matrices
\be \label{defL}
L^+_{\bf k}\equiv \g_0{\cal M}_{\bf k}^\dag {\cal M}_{\bf k} \g_0
\qquad \mbox{and} \qquad  L^-_{\bf k}\equiv {\cal M}_{\bf k}
{\cal M}_{\bf k}^\dag \,\, ,
\ee
respectively, i.e., if the number of different eigenvalues $\lambda_{k,r}$ is $n$,
the projectors are \cite{thesis},
\be \label{defP}
{\cal P}_{{\bf k},r}^\pm = \prod_{s\neq r}^n 
\frac{L_{\bf k}^\pm-\lambda_{k,s}}{\lambda_{k,r}-\lambda_{k,s}} \,\, .
\ee
In the case of spin-zero color superconductors (2SC and CFL),
$L^+_{\bf k}$ and $L^-_{\bf k}$ are identical. But this is not true for all 
spin-1 phases as we shall see in Sec.\ \ref{results}. However, in all 
phases we consider, $L_{\bf k}^+$ and $L_{\bf k}^-$ have the same spectrum.
Consequently, 
\be \label{spectrum}
L_{\bf k}^\pm = \sum_r \lambda_{k,r} {\cal P}_{{\bf k},r}^\pm \,\, .
\ee
We denote the degeneracy of the eigenvalue $\lambda_{k,r}$ by
\be \label{degeneracy}
n_r\equiv {\rm Tr}[{\cal P}_{{\bf k},r}^\pm] \,\, .
\ee
In general, the eigenvalues $\lambda_{k,r}$  
depend on the direction of the quark momentum $\uk$ (they do not depend 
on the modulus $k$). They enter the quasiparticle excitation energies 
introduced in Eq.\ (\ref{prop}),
\be \label{excite}
\e_{{\bf k},r}^e \equiv \sqrt{(ek-\m)^2+\lambda_{k,r} \, |\phi_e|^2} \,\, .
\ee
Consequently, the spectrum of the matrices $L_{\bf k}^\pm$ determines 
the structure of the quasiparticle excitations. In all phases we consider,
there are at most three different eigenvalues, and at most 
two different {\em nonzero} eigenvalues $\lambda_{k,r}$. A zero eigenvalue 
corresponds to an ungapped excitation branch. This is for instance the case in 
the 2SC phase, where the ungapped blue quarks give rise to a 
zero eigenvalue $\lambda_2=0$, while $\lambda_1=1$. Two different nonzero eigenvalues
correspond to two excitation branches with different gaps, well-known from
the CFL phase, where there is a quasiparticle singlet with gap $2\phi$
($\lambda_1=4$ with degeneracy 1) and a quasiparticle octet with gap $\phi$
($\lambda_2=1$ with degeneracy 8). In the case of the spin-1 phases, the
eigenvalues carry the potential angular dependence of the energy gap.
In the following we assume that there is no additional angular 
dependence in the functions $\phi_e$. Moreover, since we neglect the 
antiparticle gap, $\phi_-\simeq 0$, we may denote the particle gap by
$\phi\equiv \phi_+$. We assume this function to
be real, $|\phi|^2=\phi^2$, and denote the value of this function 
at the Fermi surface by $\phi_0$. With these assumptions and notations we can 
define the root-mean-square (= quadratic mean) of the function 
$\sqrt{\lambda_{k,r}}\, \phi_0$ as  
\be \label{averagegap}
\overline{\phi}_r\equiv \sqrt{\angk{\lambda_{k,r}}}\,\phi_0 \,\, .
\ee
Furthermore, for the following it is convenient to define the normalized 
eigenvalue
\be
\hat{\lambda}_{k,r} \equiv \frac{\lambda_{k,r}}{\angk{\lambda_{k,r}}} \,\, .
\ee    
We make use of these definitions in the final results, cf.\ Eqs.\ 
(\ref{gapaverage}), (\ref{ratio}) and (\ref{pressure}).

Finally, we determine the anomalous propagator $\Xi^+$. Inserting the 
expression for the propagator $G^+$, 
Eq.\ (\ref{prop}), into the definition (\ref{Xpm}) and using the form of 
the gap matrix $\Phi^+$ given in Eq.\ (\ref{gapmatrix}), we obtain 
\be \label{S212SC}
\Xi^+(K)=-\sum_{e,r} \gamma_0 \, {\cal M}_{\bf k}\, 
\gamma_0\, {\cal P}_{{\bf k},r}^+ \Lambda_{\bf k}^{-e}
\, \frac{\phi_e(K)}{\left[k_0/Z(k_0)\right]^2-
(\e_{{\bf k},r}^e)^2} \,\, . 
\ee

\subsection{Solution of the gap equation for an anisotropic gap} 
\label{general1}

In this section, we solve the gap equation for the general
case of an anisotropic gap function. Formally, with the definitions of the
previous section, this means that we allow for angular-dependent 
eigenvalues $\lambda_{k,r}$. Starting from the gap equation (\ref{gapeq21})
for the gap matrix $\Phi^+(K)$, we obtain a gap equation for the function 
$\phi_e(K)$ by inserting Eqs.\ (\ref{gapmatrix}) and (\ref{S212SC})
into Eq.\ (\ref{gapeq21}),
multiplying both sides with ${\cal M}_{\bf k}^\dag \Lambda_{\bf k}^e$, and
taking the trace over color, flavor, and Dirac space. Moreover, we use
the gluon propagator in the Hard-Dense-Loop (HDL) approximation, which is 
permissible to subleading order \cite{rischke2}. We denote quasiparticle 
energies by $\e_{{\bf q},s}\equiv \e^+_{{\bf q},s}$, because, due to 
$\phi^-\simeq 0$, the quasi{\em anti}particle energies do not occur in the gap 
equation. Since the effect of the 
wave function renormalization factor $Z(k_0)$ for the gap is known 
\cite{wang,ren}
and does not affect the angular structure of the equation, we omit it for 
simplicity. The resulting factor $b_0'$, see Eq.\ (\ref{tildeb}) can easily 
be reinserted into the final result. We arrive at
\be
\phi(K) = g^2 \frac{T}{V}\sum_Q\sum_s\,\frac{\phi(Q)}{q_0^2-\e_{{\bf q},s}^2}
\,\D^{\m\n}(K-Q)\,{\cal T}_{\m\n}^s({\bf k},{\bf q}) \,\, ,
\ee
where, following Ref.\ \cite{schmitt}, we have defined
\be \label{defT}
{\cal T}_{\m\n}^s({\bf k},{\bf q}) \equiv -\frac{{\rm Tr}\left[\g_\m\, T^T_a\,
\g_0\,{\cal M}_{\bf q}\,\g_0 \,{\cal P}_{{\bf q},s}^+\,\Lambda_{\bf q}^-\,
\g_\n\, T_a\,{\cal M}_{\bf k}^\dag\,\Lambda_{\bf k}^+\right]}{{\rm Tr}\left[
{\cal M}_{\bf k}\,{\cal M}_{\bf k}^\dag\,\Lambda_{\bf k}^+\right]} \,\, .
\ee
With $P\equiv K-Q$, the gluon propagator in pure Coulomb gauge is
given by 
\be
\Delta^{00}(P)=\Delta_\ell(P) \,\, , \,\,\, \Delta^{0i}(P)=0 \,\, ,
\,\,\,  \Delta^{ij}(P)=(\delta^{ij}-\hat{p}^i\hat{p}^j)\, \Delta_t(P)
\,\, .
\ee
For the definition of the longitudinal and transverse gluon propagators 
$\Delta_{\ell,t}$ see for instance Ref.\ \cite{rischke}. 
With the (negative) transverse projection of the tensor 
${\cal T}_{\m\n}^s({\bf k},{\bf q})$, 
\be \label{Ttrans}
{\cal T}_t^s({\bf k},{\bf q})\equiv -(\delta^{ij}-\hat{p}^i\hat{p}^j) 
\, {\cal T}_{ij}^s({\bf k},{\bf q}) \,\, ,
\ee
we obtain after performing the Matsubara sum (for details see 
Ref.\ \cite{rischke}), and after taking the thermodynamic limit, $V\to \infty$,
\be \label{matsu}
\phi_{k,r} = \frac{g^2}{4}\,\int\frac{d^3q}{(2\pi)^3}\,\sum_s\,
\frac{\phi_{q,s}}{\e_{{\bf q},s}}\,\tanh\left(\frac{\e_{{\bf q},s}}{2T}\right)
\Big[F_\ell(p)\,{\cal T}_{00}^s({\bf k},{\bf q}) + 
F_t(p,\e_{{\bf q},s},\e_{{\bf k},r})\,
{\cal T}_t^s({\bf k},{\bf q})\Big] \,\, ,
\ee
where 
\be
F_\ell(p)\equiv \frac{2}{p^2 + 3m_g^2}
\ee
arises from static electric gluons, while
\be \label{Ft}
F_t(p,\e_{{\bf q},s},\e_{{\bf k},r}) \equiv \frac{2}{p^2}\, \Theta(p-M)
+\Theta(M-p)\left[\frac{p^4}{p^6+M^4(\e_{{\bf q},s}+\e_{{\bf k},r})^2}
+\frac{p^4}{p^6+M^4(\e_{{\bf q},s}-\e_{{\bf k},r})^2}\right]
\ee
originates from non-static and almost static magnetic gluons. In 
Eq.\ (\ref{matsu}) we have abbreviated 
$\phi_{k,r}\equiv \phi(\e_{{\bf k},r},{\bf k})$ and 
$\phi_{q,s}\equiv \phi(\e_{{\bf q},s},{\bf q})$.
At this point, the angular integral $d\Omega_q$ on the right-hand
side of Eq.\ (\ref{matsu}) seems to be too complicated, since, besides the 
square of the gluon 3-momentum 
$p^2=k^2+q^2-2\,{\bf q}\cdot{\bf k}$ and the functions ${\cal T}_{00}^s$, 
${\cal T}_t^s$, also the excitation energies $\e_{{\bf q},s}$ depend
on the direction of ${\bf q}$. The solution to this problem is to multiply 
both sides of the equation with 
\be \label{ell}
\ell_k\equiv {\rm Tr}\left[
{\cal M}_{\bf k}\,{\cal M}_{\bf k}^\dag\,\Lambda_{\bf k}^+\right]=
\frac{1}{2}\sum_r n_r\lambda_{k,r} \,\, ,
\ee 
and take the angular average over $\uk$ of the whole equation. 
The right-hand side of Eq.\ (\ref{ell}) is obtained with the help of 
Eqs.\ (\ref{defL}), (\ref{spectrum}), (\ref{degeneracy}), and the following identities,
\be \label{traces}
\frac{1}{2}\Tr[{\cal P}_{{\bf k},r}^+] 
=\Tr[{\cal P}_{{\bf k},r}^+\Lambda_{\bf k}^e]
=\Tr[{\cal P}_{{\bf k},r}^-\Lambda_{\bf k}^e] \,\, .
\ee
They will become obvious in Sec.\ \ref{results}, where we discuss the
specific phases in detail. The only nontrivial phase with respect to 
these identities is the A phase, where 
${\cal P}_{{\bf k},r}^+ \neq {\cal P}_{{\bf k},r}^-$. In this case, one uses 
Eqs.\ (\ref{pA}) from Appendix \ref{AppA} to prove these relations. 

We arrive at
\be
\left\langle\ell_k\right\rangle_{\uk}\,\phi_{k,r}
= \frac{g^2}{4}\,\int\frac{d^3q}{(2\pi)^3}\,\sum_s\,
\frac{\phi_{q,s}\;\ell_q}{\e_{{\bf q},s}}\,
\tanh\left(\frac{\e_{{\bf q},s}}{2T}\right)
 \;
\left\langle F_\ell(p)\,\frac{\ell_k}{\ell_q}\,
{\cal T}_{00}^s({\bf k},{\bf q}) + 
F_t(p,\e_{{\bf q},s},\e_{{\bf k},r})\,\frac{\ell_k}
{\ell_q}\,{\cal T}_t^s({\bf k},{\bf q})\right\rangle_{\uk} \,\, ,
\label{pullout}
\ee
where we pulled the factor $\ell_q$ out of the $\uk$ integral on
the right-hand side of the equation.
This turns out to be convenient for the calculation. On the left-hand
side we have made use of the assumption that the function $\phi_{k,r}$
does not depend on the direction of ${\bf k}$. We shall see below that 
this assumption is consistent with our final result.

Now the angular
integral over $\uk$ has to be performed for each phase separately. 
In Appendix \ref{transversephases} we present this calculation explicitly for 
the transverse polar, planar, and A phases, and in Appendix \ref{AppB}
it is presented for arbitrary longitudinal phases. 
However, we can give a general result similar to the method 
introduced in Ref.\ \cite{schmitt}.  
For the angular integral we use a frame with $z'$-axis parallel to 
${\bf q}$. However, besides the fixed direction given by ${\bf q}$, in all 
nontrivial cases there is 
another preferred direction in the system, namely the one that is 
picked by the order parameter. This has to be taken into account for the
integral and is explained in detail  
in Appendices \ref{transversephases} and \ref{AppB}. We neglect the 
$\uk$-dependence in $\e_{{\bf k},r}$ occurring in the function $F_t$. 
Within this approximation, the functions $F_\ell$, $F_t$ do not depend
on the azimuthal angle $\varphi'$, they only enter the $\uk$ integral 
through the polar angle $\theta'$. The $\theta'$ integral can be transformed 
into an integral over $p$. 
This $p$ integral is performed using the formalism presented in 
Ref.\ \cite{schmitt}. To subleading order, we may set $k\simeq q\simeq\mu$, 
which allows us, for all phases we consider, to write the result of the 
azimuthal integral in the following power series of $p$,
\begin{subequations} \label{eta}
\bea
\frac{1}{2\pi}\int_0^{2\pi}d\varphi'\;\frac{\ell_k}
{\ell_q} \;{\cal T}^s_{00}({\bf k},{\bf q}) &\simeq& 
a_s\left(\eta_0^\ell + \eta_2^\ell\,
\frac{p^2}{\mu^2} + \eta_4^\ell\,\frac{p^4}{\mu^4}\right) \,\, , \\
\frac{1}{2\pi}\int_0^{2\pi}d\varphi'\;\frac{\ell_k}
{\ell_q} \;{\cal T}^s_t({\bf k},{\bf q}) &\simeq & a_s
\left(\eta_0^t + \eta_2^t\,\frac{p^2}{\mu^2} + 
\eta_4^t\,\frac{p^4}{\mu^4}\right) \,\, . 
\eea
\end{subequations}
The coefficients $a_s$ are chosen such that they add up to one, cf.\ 
Eq.\ (\ref{condition}). In all cases we consider, we find
\be \label{asgeneral}
a_s = \frac{n_s\lambda_{q,s}}{\sum_r n_r\lambda_{q,r}} \,\, ,
\ee
and there are at most 
two nonzero coefficients $a_1$, $a_2$. 
In general, the coefficients $a_s$ and $\eta^{\ell,t}$ depend on the polar 
angle $\theta$ of the vector $\hat{\bf q}$ 
($\cos\theta = \hat{\bf q}\cdot{\bf e}_z$). However, in most of the cases 
we consider, they are constant. The only phase with angular-dependent coefficients
$\eta^{\ell,t}$ is the mixed polar phase, while the only phase with angular-dependent 
coefficients $a_s$ is the transverse A phase.    

With $\eta_0^\ell=\eta_0^t=2/3$, which holds for all phases we consider, 
we arrive at (for details of the $p$ integral, see Ref.\ \cite{schmitt})
\be \label{gapeq3}
\left\langle\ell_k\right\rangle_{\uk}\;\phi_{k,r}
=\int \frac{d\Omega_q}{4\pi}\;\ell_q\;\bar{g}^2
\int_0^\d d(q-\mu)\;\sum_s\,a_s\,\frac{\phi_{q,s}}{\e_{{\bf q},s}}
\;\frac{1}{2}\,\ln\left(\frac{\tilde{b}^2\mu^2 e^{-2d}}
{|\e_{{\bf q},s}^2- \e_{{\bf k},r}^2|}
\right)\,\, ,
\ee
with $\tilde{b}$ as defined in Eq.\ (\ref{tildeb}). The value of $d$ 
can be determined from the coefficients in Eqs.\ (\ref{eta}), 
\be \label{d}
d=-\frac{6}{\eta_0^t}\, \left[\eta_2^\ell+\eta_2^t+2(\eta_4^\ell+\eta_4^t)
\right] \,\, .
\ee
Consequently, the result of the $\uk$ integral in the gap equation can be
obtained by computing the coefficients of the power series in $p$, since
the $p$ integral is generic for all phases. This result
is very similar to that of Ref.\ \cite{schmitt}, where the ${\bf q}$
integral on the right-hand side of the gap equation could be performed 
directly because of the trivial angular structure. In the present, more 
general, case, however, the ${\bf q}$ integral still has to be done. In 
Eq.\ (\ref{gapeq3}), this integral has been divided into its angular part and 
the integral over the modulus $q$, which can be restricted to an integral 
over a small 
region of size $2\d$ around the Fermi surface, where $\d$ is much larger than 
the gap but much smaller than the chemical potential. This integral can now
be done in the usual way, keeping the angular 
dependence in the excitation energy $\e_{{\bf q},s}$. We briefly
repeat this calculation, which leads to Eq.\ (\ref{finalgapeq}), 
more details for the case of two different 
gaps can be found in Ref.\ \cite{schmitt}. With the approximation \cite{son}
\be
\frac{1}{2}\,\ln\left(\frac{\tilde{b}^2\m^2e^{-2d}}{|\e_{{\bf q},s}^2-
\e_{{\bf k},r}^2|}\right)
\simeq\Theta(\e_{{\bf q},s}-\e_{{\bf k},r})\, \ln 
\left(\frac{\tilde{b}\,\m\,e^{-d}}{\e_{{\bf q},s}}\right)
+\Theta(\e_{{\bf k},r}-\e_{{\bf q},s})\, \ln \left(\frac{\tilde{b}\,\m\,e^{-d}}
{\e_{{\bf k},r}}\right) 
\ee
and the new variables
\be \label{vartrans}
x_r \equiv \bar{g} \, \ln\left(\frac{2\tilde{b}\,\m\,e^{-d}}{k-\m+\e_{{\bf k},r}}\right) \quad, \qquad
y_s \equiv \bar{g} \, \ln\left(\frac{2\tilde{b}\,\m\,e^{-d}}{q-\m+\e_{{\bf q},s}}\right) \,\, ,
\ee
to subleading order the gap equation (\ref{gapeq3}) transforms into    
\bea \label{gapeqnew}
\left\langle\ell_k\right\rangle_{\uk}\;\phi(x_r)
&=&\int \frac{d\Omega_q}{4\pi}\;\ell_q\;
\sum_s\,a_s\,\left\{x_r\;\int_{x_r}^{x_s^*} dy_s \, \tanh\left[\frac{\e(y_s)}
{2T}\right]\;\phi(y_s) \right.\nonumber \\ 
&&\left. + \int_{x_0}^{x_r} dy_s \,y_s \, \tanh\left[
\frac{\e(y_s)}{2T}\right] \;\phi(y_s) \right\}\,\, ,
\eea
where we defined
\be \label{xs}
x_s^*\equiv \bar{g}\,\ln\left(\frac{2\tilde{b}\,\m\,e^{-d}}
{\sqrt{\lambda_{q,s}}\,\phi_0}\right)\,\, , 
\qquad x_0\equiv \bar{g}\,\ln\left(\frac{\tilde{b}\,\m\,e^{-d}}{\d}
\right) \,\, ,
\ee
with $\phi_0\equiv \phi(x_1^*)\simeq \phi(x_2^*)$. In the case of an 
isotropic gap, $\phi_0$ is the value of the gap at the Fermi surface.
Note that Eq.\ (\ref{gapeqnew}) corresponds to Eq.\ (79) of Ref.\ \cite{schmitt}.
The difference to that equation, besides the angular integral, 
arises from the simplification we made above by omitting the wave function 
renormalization factor. 

While Eq.\ (\ref{gapeq3})
in principle is a set of two equations, one for each excitation branch, 
labelled by the index $r$, we see from Eq.\ (\ref{gapeqnew}), that, after
the change of variables and neglecting sub-subleading contributions, 
one equation for the variable $x_r$ is left. We can write
this single equation for the renamed variable $x$ in the form    
\begin{eqnarray} 
\left\langle\ell_k\right\rangle_{\uk}\;
\phi(x) & = & \int \frac{d\Omega_q}{4\pi}\;\ell_q 
\left\{ x \int_x^{x_2^*}dy\, 
\tanh\left[\frac{\e(y)}{2T}\right]  \, \phi(y)
+ \int_{x_0}^x dy\, y\,  
\tanh \left[\frac{\e(y)}{2T}\right]\, \phi(y) \right.\nonumber \\
&  &\left. - a_1\, x \int_{x_1^*}^{x_2^*}dy\, 
\tanh\left[\frac{\e(y)}{2T}\right]\, \phi(y)\right\} \,\, .
\label{phix}
\end{eqnarray}
The solution of this gap equation for $T=0$ is found in the usual way. 
Differentiating twice with respect to $x$ yields a second-order differential 
equation. 
The two constants of the general solution are determined with the help of
Eq.\ (\ref{phix}) and its first derivative at the point $x=x_2^*$.
One obtains
\be \label{solution}
\phi(x) = \phi_0\,\left[\cos(x_2^*-x)+a_1\,(x_2^*-x_1^*)\,\sin(x_2^*-x)\right]
\,\, .
\ee
An exchange of the indices 1 and 2 in this solution yields
the same final result for $\phi_0$. In order to determine $\phi_0$, one 
inserts the solution (\ref{solution}) into Eq.\ (\ref{phix}) and 
evaluates the equation at the point $x=x_2^*$. Then, the integrals on the 
right-hand side
of the equation are trivial. Using $\sin(\a-\b)=\sin\a\,\cos\b - \cos\a\sin\b$
and $\cos(\a-\b) = \cos\a\,\cos\b + \sin\a\,\sin\b$ and the 
approximations $\sin x_0\simeq x_0$, $\cos x_0 \simeq 1$ (note that 
$x_0$ is parametrically of order $O(\bar{g})$), we obtain 
\be \label{finalgapeq}
0 = \int\frac{d\Omega_q}{4\pi}\, \ell_q\, \left[\cos x_2^*
+ a_1\,(x_2^*-x_1^*)\;\sin x_2^*\right] \,\, ,
\ee
where $\angk{\ell_k}=\angq{\ell_q}$ has been 
subtracted on both sides of the equation.
With the definition for $x_2^*$ in Eq.\ (\ref{xs}) and the approximations
\begin{subequations}
\bea
\cos x_2^*
&\simeq& \cos\left[\bar{g}
\,\ln\left(\frac{2\tilde{b}\m}{\phi_0}\right)\right] + \bar{g}\,
(\ln\sqrt{\lambda_{q,2}} + d)\; 
\sin\left[\bar{g}
\,\ln\left(\frac{2\tilde{b}\m}{\phi_0}\right)\right] \,\, , \\
\sin x_2^*
&\simeq&\sin\left[\bar{g}
\,\ln\left(\frac{2\tilde{b}\m}{\phi_0}\right)\right] -\bar{g}\,
(\ln\sqrt{\lambda_{q,2}} + d)\;
\cos\left[\bar{g}
\,\ln\left(\frac{2\tilde{b}\m}{\phi_0}\right)\right] \,\, ,
\eea
\end{subequations}
we find (using $a_1 + a_2 = 1$)
\be \label{almostdone}
0 = \cos\left[\bar{g}
\,\ln\left(\frac{2\tilde{b}\m}{\phi_0}\right)\right] + \bar{g}\,
(\overline{\zeta}
+ \overline{d}) \;
\sin\left[\bar{g}
\,\ln\left(\frac{2\tilde{b}\m}{\phi_0}\right)\right] \,\, .
\ee
We use the abbreviations  
\be \label{zeta}
\overline{\zeta}\equiv \frac{\left\langle\ell_q\,\zeta\right\rangle_{\uq}}
{\left\langle \ell_q\right\rangle_{\uq}} 
\,\, , \qquad \overline{d}\equiv 
\frac{\left\langle\ell_q\,d\right\rangle_{\uq}}
{\left\langle \ell_q\right\rangle_{\uq}} \,\, ,
\ee
with $\zeta=\ln(\lambda_{q,1}^{a_1}\lambda_{q,2}^{a_2})^{1/2}$, as defined in 
Eq.\ (\ref{zetaiso}) for isotropic gaps. With Eqs.\ (\ref{ell}) and (\ref{asgeneral}),
we can write $\overline{\zeta}$ in the following simple form,
\be \label{zetafromL}
\overline{\zeta} = \frac{1}{2}
\frac{\langle n_1 \lambda_{q,1}\ln\lambda_{q,1} + n_2 \lambda_{q,2}\ln\lambda_{q,2}
\rangle_{\uq}}{\langle n_1 \lambda_{q,1} +  n_2 \lambda_{q,2}\rangle_{\uq}} \,\, .
\ee
This expression shows that $\overline{\zeta}$ can be determined solely from the 
spectrum of the matrix $L^+_{\bf q}$, cf.\ Eqs.\ (\ref{defL}), (\ref{spectrum}), and
(\ref{degeneracy}).

From Eq.\ (\ref{almostdone}) we deduce that, to subleading order 
(reinserting the factor $b_0'$),
\be \label{gapgeneral}
\phi_0 = 2\,\tilde{b}\,b_0'\,e^{-\overline{d}}\,e^{-\overline{\zeta}} \,\mu \, 
\exp\left(-\frac{\pi}{2\bar{g}}\right) \,\, .
\ee
This result, which is the main result of this paper, is the
generalization of Eq.\ (\ref{gapconstant}). It says that, in the case 
of anisotropic gaps and/or an angular-dependent value of $d$, one has to 
replace the exponents $d\to\overline{d}$, $\zeta\to\overline{\zeta}$. 
It turns out that in most of the cases we consider, $d$ is a constant number.
Only in the mixed polar phase, see Sec.\ \ref{results}, $\overline{d}\neq d$.
However, the modification of the other exponent, $\overline{\zeta}$, plays 
an important and nontrivial role, especially for the determination 
of the ground state. 
 
In order to compare the result with the corresponding one in the theory 
of superfluid $^3$He, we consider the special case of only one gapped
(but anisotropic) excitation branch, $\lambda_{q,2}=a_2=0$. In this case,
\be \label{onegap}
e^{-\overline{\zeta}} \to \frac{1}{\sqrt{\angq{\lambda_{q,1}}}}\,
\exp\left(-\frac{1}{2}
\angq{\hat{\lambda}_{q,1} \ln\hat{\lambda}_{q,1}}\right) \,\, ,
\ee
and the quadratic mean of the gap is given by
\be \label{gapaverage} 
\overline{\phi}_1 = 2\,\tilde{b}\,b_0'\,e^{-\overline{d}} \,\mu \, 
\exp\left(-\frac{\pi}{2\bar{g}}\right)
\,\exp\left(-\frac{1}{2}
\angq{\hat{\lambda}_{q,1} \ln\hat{\lambda}_{q,1}}\right) \,\, .
\ee
In this special case, the exponent involving the angular dependence of the 
gap, is exactly the same as in $^3$He, cf.\ Eq.\ (3.63) of 
Ref.\ \cite{vollhardt}.

\subsection{The critical temperature} \label{general2}

In order to determine the critical temperature $T_c$, we proceed similar to 
the above calculation of the gap: Starting from the gap equation 
(\ref{phix}), we apply the method presented in Ref.\ \cite{schmitt}
before doing the $d\Omega_q$-integral. The basic assumption is that 
the shape of the gap function does not change with temperature, i.e., 
we employ the following factorization of the temperature-dependent
gap function,
\be
\phi(x,T) \simeq \phi(T)\, \frac{\phi(x,0)}{\phi_0} \;,
\ee
where $\phi(T)\equiv\phi(x_2^*,T)$ is the value of the gap 
at the Fermi surface at temperature $T$,
and $\phi(x,0)$ is the zero-temperature gap function $\phi(x)$ computed in the
last section. This ansatz, inserted into Eq.\ 
(\ref{phix}), yields at the Fermi surface
\begin{eqnarray} \label{tempeq} 
\left\langle\ell_k\right\rangle_{\uk} 
& = & \int \frac{d\Omega_q}{4\pi}\;\ell_q 
\left\{ \int_{x_0}^{x_\kappa}dy\,y\, 
\tanh\left[\frac{\e(y)}{2T}\right]  \, \frac{\phi(y,0)}{\phi_0}
+ \int_{x_\kappa}^{x_2^*} dy\, y\,  
\tanh \left[\frac{\e(y)}{2T}\right]\, \frac{\phi(y,0)}{\phi_0} 
\right.\nonumber \\
&  &\left. - a_1\, x \int_{x_1^*}^{x_2^*}dy\, 
\tanh\left[\frac{\e(y)}{2T}\right]\, \frac{\phi(y,0)}{\phi_0}\right\} \,\, ,
\end{eqnarray}
where the second integral in Eq.\ (\ref{phix}) has been divided into two 
integrals: One running from $x_0$ to $x_\kappa$, with 
$x_\kappa\equiv x_2^*-\bar{g}\,\ln(2\kappa)$, $\kappa\gg 1$, and one running 
from $x_\kappa$ to $x_2^*$. In Ref.\ \cite{schmitt} it is shown how the 
integrals in the angular brackets on the right-hand side of Eq.\ (\ref{tempeq})
are evaluated. The result is 
\be
0 = \int \frac{d\Omega_q}{4\pi}\;\ell_q\,\ln\left(\frac{e^\g\phi_0}
{\pi T_c}\sqrt{\lambda_{q,1}^{a_1}\,\lambda_{q,2}^{a_2}}\right) \,\, .
\ee
This leads to 
\be \label{nonbcs}
\frac{T_c}{\phi_0} = \frac{e^\g}{\pi}\; e^{\overline{\zeta}} \,\, .
\ee
Note that, although the ratio $T_c/\phi_0$ depends on the constant 
$\overline{\zeta}$,
the absolute value for $T_c$ does not, since the factor 
$e^{-\overline{\zeta}}$ of the gap $\phi_0$ cancels the factor 
$e^{\overline{\zeta}}$ on the right-hand side 
of Eq.\ (\ref{nonbcs}). Consequently, in units of the critical temperature
in the 2SC phase, $T_c^{\rm 2SC}$, the critical temperature only depends 
on the constant $\overline{d}$, $T_c/T_c^{\rm 2SC} = e^{-\overline{d}}$. 

We recover as a special case the result for two isotropic gaps, 
Eq.\ (\ref{tempconstant}). For the case of only one gapped excitation 
branch, and using Eq.\ (\ref{onegap}), the ratio between the critical
temperature and the quadratic mean of the gap at the Fermi surface for 
$T=0$ reads
\be \label{ratio}
\frac{T_c}{\overline{\phi}_1} = \frac{e^\g}{\pi}\;\exp\left(\frac{1}{2}\,
\angq{\hat{\lambda}_{q,1}\,\ln\hat{\lambda}_{q,1}}\right) \,\, .
\ee  
Consequently, we find a nontrivial modification of the BCS relation
$T_c=0.57\phi_0$ also in the case of a single, but anisotropic gap.
Obviously, the BCS relation is recovered from Eq.\ (\ref{ratio}) for
the case of a constant gap, since in this case $\hat{\lambda}_{q,1}=1$. 
Note that the result (\ref{ratio}) 
is identical to the one in the theory of superfluid 
$^3$He, cf.\ Eq.\ (3.63) of Ref.\ \cite{vollhardt}.

\subsection{The pressure}  \label{general3}

The starting point for the calculation of the pressure is the effective
action, Eq.\ (\ref{Gamma1}). In the two-loop approximation, we can write 
\be
\G_2[\Delta,{\cal S}] = \frac{1}{4}\Tr(\Sigma {\cal S}) \,\, .
\ee
Then, making use of the Dyson-Schwinger equation (\ref{DS2}), the fermionic 
part of the effective potential at the stationary point can be written as
\be \label{potgeneral}
\G[{\cal S}] = \frac{1}{2}\Tr\ln {\cal S}^{-1} - 
\frac{1}{4}\Tr\left(1-{\cal S}_0^{-1}{\cal S}\right) 
\,\, .
\ee
In order to evaluate the first term on the right-hand side of this 
equation, we use the identity $\Tr\ln\,{\cal S}^{-1} = 
\ln{\rm det}\,{\cal S}^{-1}$ and the fact that for 
arbitrary matrices 
${\cal A}$, ${\cal B}$, ${\cal C}$, and an invertible matrix ${\cal D}$, 
\be  
{\rm det} \left(\begin{array}{cc}
{\cal A} & {\cal B} \\ {\cal C} & {\cal D} \end{array}\right)
 = {\rm det} ({\cal A}\,{\cal D} -{\cal B}\,{\cal D}^{-1}{\cal C}\,{\cal D})
\,\, .
\ee 
Then, making use of ${\cal S}^{-1} = {\cal S}_0^{-1} + \Sigma$ and Eqs.\ 
(\ref{free}) and (\ref{sigmanambu}), the trace over Nambu-Gor'kov space yields 
\be 
\frac{1}{2}\Tr\ln {\cal S}^{-1}= \frac{1}{2}\Tr\ln \left\{
([G_0^+]^{-1} + \Sigma^+)\,
([G_0^-]^{-1} + \Sigma^-) - 
\Phi^-\,([G_0^-]^{-1} + \Sigma^-)^{-1}\, 
\Phi^+\,([G_0^-]^{-1} + \Sigma^-)\right\} \,\, .
\ee
In order to proceed, we set $Z(k_0)\simeq 1$ in the fermion propagator 
$G^{\pm}$, cf.\ its definition in Eq.\ (\ref{prop}). Using the identity 
\be \label{help}
[G_0^\mp]^{-1}[G_0^\pm]^{-1} = \sum_e[k_0^2-
(\m-e\, k)^2]\Lambda_{\bf k}^{\pm e} \,\, ,
\ee
we find 
\bea 
\frac{1}{2}\Tr\ln {\cal S}^{-1} &=& \frac{1}{2}\Tr\ln 
\sum_e\left[k_0^2-(\m-ek)^2
-\phi_e^2 L_{\bf k}^+\right] 
\Lambda_{\bf k}^{-e}   \nonumber  \\
&=& \frac{1}{2}\,\sum_{e,r} \sum_K \Tr[{\cal P}_{{\bf k},r}^+\Lambda_{\bf k}^{-e}]
\,\ln\left[k_0^2 - (\e_{{\bf k},r}^e)^2\right] \,\, .
\eea
After performing the Matsubara sum, this expression reads
\be
\label{firstterm}
\frac{1}{2}\Tr\ln {\cal S}^{-1} =  \frac{1}{2}\frac{V}{T}\sum_{e,r}
\int\frac{d^3 k}{(2\pi)^3}
\Tr[{\cal P}_{{\bf k},r}^+\Lambda_{\bf k}^{-e}]
\left\{\e_{{\bf k},r}^e + 
2\,T\,\ln\left[1 + \exp\left(-\frac{\e_{{\bf k},r}^e}{T}\right)
\right]\right\}
\,\, ,
\ee
where the trace now runs only over color, flavor, and 
Dirac space.

The Nambu-Gor'kov trace for the second term on the right-hand side of 
Eq.\ (\ref{potgeneral}) is easily performed with the definitions (\ref{free})
and (\ref{fullquark}). Note that the anomalous propagators $\Xi^\pm$, 
occuring in the full quark propagator ${\cal S}$, do not enter the result. 
One obtains
\be 
\frac{1}{4}\Tr(1-{\cal S}_0^{-1}{\cal S}) = 
-\frac{1}{4}\sum_{e,r}\sum_K\, \Tr[{\cal P}_{{\bf k},r}^+
\Lambda_{\bf k}^e + {\cal P}_{{\bf k},r}^-\Lambda_{\bf k}^{-e}]\, 
\frac{\lambda_{k,r}\,\phi_e^2}{k_0^2-(\e_{{\bf k},r}^e)^2} \,\, , 
\ee
which becomes, after performing the Matsubara sum,
\be \label{secondterm}
\frac{1}{4}\Tr(1-{\cal S}_0^{-1}{\cal S})= \frac{1}{4}\frac{V}{T}
\sum_{e,r}\int\frac{d^3 k}{(2\pi)^3}\,\Tr[{\cal P}_{{\bf k},r}^+
\Lambda_{\bf k}^e + {\cal P}_{{\bf k},r}^-\Lambda_{\bf k}^{-e}] 
\; \frac{\lambda_{k,r}\,
\phi_e^2(\e_{{\bf k},r}^e,k)}{2\,\e_{{\bf k},r}}
\,\tanh\frac{\e_{{\bf k},r}^e}{2\,T} \,\, .
\ee
Using Eq.\ (\ref{traces}), the final result for the pressure 
$p=-V_{\rm eff}=\frac{T}{V}\G$, 
obtained by putting together Eqs.\ (\ref{firstterm}) and (\ref{secondterm}), is
\be 
p = \frac{1}{4}\sum_{e,r}\int\frac{d^3 k}{(2\pi)^3} \,
\Tr[{\cal P}_{{\bf k},r}^+] \,  
\left\{\e_{{\bf k},r}^e + 
2\,T\,\ln\left[1 + \exp\left(-\frac{\e_{{\bf k},r}^e}{T}\right)
\right] -  \frac{\lambda_{k,r}\,
\phi_e^2(\e_{{\bf k},r}^e,k)}{2\,\e^e_{{\bf k},r}}
\,\tanh\frac{\e_{{\bf k},r}^e}{2\,T}
\right\} \,\, .\label{Tnonzero}
\ee
In the following, we restrict ourselves to the zero-temperature case, $T=0$. 
Furthermore, we neglect the antiparticle gap and thus 
denote $\phi\equiv\phi_+$, as above. In this case, Eq.\ (\ref{Tnonzero}) 
becomes
\be
p = \frac{1}{4}\sum_{r}\int\frac{d^3 k}{(2\pi)^3} \,
n_r\,  
\left(\e_{{\bf k},r}^+ + \e_{{\bf k},r}^- 
 - \frac{\lambda_{k,r}\, \phi^2_{k,r}}{2\,\e^+_{{\bf k},r}}
\right) \,\, .
\ee
In order to evaluate the integral over the absolute 
value of the quark momentum, we assume the gap function to be constant
in a small region around the
Fermi surface of size $2\delta$, $\phi_{k,r}=\phi_0$, and zero elsewhere.
Moreover, we use the integrals 
\be
\int_0^\d d\xi \, \left(\sqrt{\xi^2+\phi_0^2}-\frac{1}{2}
\frac{\phi_0^2}{\sqrt{\xi^2+\phi_0^2}}\right) = \frac{1}{2}\,\d\,
\sqrt{\d^2+\phi_0^2}
= \frac{1}{2}\d^2 + \frac{1}{4} \phi_0^2  + O\left(\frac{\phi_0^4}{\d^2}\right)
\,\, ,
\ee
and
\be
\int_0^\infty dk\,k^2(\e_{k0}^+ + \e_{k0}^- - 2k)  = \frac{1}{6}\mu^4 \,\, ,
\ee
where the vacuum energy $2k$ has been subtracted and $\e_{k0}^e\equiv|k-e\mu|$.
Neglecting terms of order $O(\phi_0^4)$, we find
\be \label{totalpressure}
p = \frac{\mu^4}{48\pi^2}\sum_r n_r + \Delta p \,\, .
\ee
where we denote the difference of the superconducting phase to that of the 
normal phase by 
\be \label{pressure}
\Delta p = \frac{\m^2}{16\,\pi^2}\,\sum_r n_r
\overline{\phi}_r^2 \,\, .
\ee
As expected from physical intuition and from the theory of $^3$He, cf.\ Eq.\ 
(\ref{condenergy}), the favored phase is determined by the largest 
condensation energy $\Delta p$, which is proportional to the sum of the 
angular averages
of the squares of the gap, weighted with the corresponding degeneracies $n_r$
of the excitation branches. (Note that, since $\overline{\phi}_r=
\sqrt{\langle\lambda_{k,r}\rangle_{\uk}}\,\phi_0$ is the {\em quadratic mean} of the 
gap, $\overline{\phi}_r^2=\langle\lambda_{k,r}\rangle_{\uk}\,\phi_0^2$ is the 
angular average of the square of the gap, not the square of the angular 
average, as the notation might suggest.)

As a special case, one finds for the 2SC phase, counting Dirac, color, and 
flavor degrees of freedom, 
\be
\Delta p_{\rm 2SC} = \frac{\m^2 (\phi_0^{\rm 2SC})^2\,N_f}{2\,\pi^2} \,\, .
\ee
This is in accordance with Ref.\ \cite{miransky}, when one identifies the 
pressure with the negative value of the effective potential at the global 
minimum.

\section{Discussion of the polar, planar, A, and CSL phases} \label{results}

In this section, we use the general results of the previous sections
for a discussion of the physical properties of certain phases in a spin-one 
color superconductor. In particular, we determine the ground state at
zero temperature.  
We focus
on the four ``inert'' phases presented in Sec.\ \ref{patterns}, the
polar, planar, A, and CSL phases, see Fig.\ \ref{symmetries}.

The common structure of all spin-one phases is given by the matrix 
${\cal M}_{\bf k}$, that determines the color and Dirac structure of
the gap matrix $\Phi^+(K)$, Eq.\ (\ref{gapmatrix}). The 
most general form of this matrix has been introduced in Eq.\ (\ref{Mgeneral}).
Specifying the spin-triplet structure $\kappa_j$, we write the matrix
as 
\be \label{Mk}
{\cal M}_{\bf k} = \sum_{i,j=1}^3 J_i\, \D_{ij}\left[\a\,\hat{k}_j + 
\b\,\g_{\perp,j}({\bf k})\right] \,\, .
\ee
The first term in angular brackets, proportional to the $j$-th component
of the fermion momentum unit vector $\hat{k}_j$, describes pairing of quarks 
with the same chirality, since it commutes with the chirality projector 
${\cal P}_{r,\ell}=(1\pm\g_5)/2$. The second one, proportional to
\be 
\g_{\perp,j}(\uk)\equiv \g_j - \hat{k}_j\vg\cdot\uk \,\, ,\qquad j=1,2,3
\,\, ,
\ee
corresponds to pairing of quarks of opposite chirality, since commuting
this term with the chirality projector flips the sign of chirality. 
In the above ansatz for the gap matrix we allow for a general linear 
combination 
of these two terms, determined by the real coefficients $\a$ and $\b$ with 
\be \label{normalize}
\a^2 + \b^2 = 1 \,\, .
\ee
In Refs.\ \cite{rischke,schmitt} the special cases $(\a,\b)=(1,0)$ and 
$(\a,\b)=(0,1)$ were 
termed ``longitudinal'' and ``transverse'' gaps, respectively. We 
shall also use these terms in the following. (In Ref.\ \cite{schaefer},
the LL and RR gaps correspond to the longitudinal and the LR and RL gaps
to the transverse gaps.) The reason why both cases can be studied 
separately is that a purely 
longitudinal gap matrix on the right-hand side of the gap equation 
does not induce a transverse gap on the left-hand side and
vice versa. More precisely, inserting the matrix ${\cal M}_{\bf k}$ from
Eq.\ (\ref{Mk}) with $\b=0$ into the anomalous propagator from Eq.\
(\ref{S212SC}), and the result into the right-hand side of the gap equation 
(\ref{gapeq21}), we realize that the Dirac structure still commutes with 
$\g_5$ and thus preserves chirality. The analogous argument holds
for the transverse gap, $\a=0$.
For the case of an equal admixture of longitudinal and transverse gaps, i.e., 
$\a=\b=1/\sqrt{2}$, let us use the term ``mixed'' gap. For the
physical results, we focus exclusively on the longitudinal, mixed, and 
transverse gaps.

\begin{table}  
\begin{tabular}[t]{|c||c|c|c|}
\hline
 & \multicolumn{3}{c}{\bf Polar phase} \vline \\ \hline
 & \multicolumn{3}{c}{${\cal M}_{\bf k} = J_3\,[\a\,\hat{k}_3 + 
\b\,\g_{\perp,3}(\uk)]$}\vline \\ \hline
 & \multicolumn{3}{c}{$L_{\bf k}^\pm = J_3^2\left[\b^2+(\a^2-\b^2)\,
\cos^2\theta\right]$}\vline \\ \hline
 & \multicolumn{3}{c}{$\lambda_{k,1} = \b^2 + (\a^2-\b^2)\,\cos^2\theta \qquad 
(n_1=8)\quad, \qquad \lambda_{k,2} = 0 \qquad (n_2=4)$}\vline \\ 
\hline
 & \multicolumn{3}{c}{${\cal P}_{{\bf k},1}^\pm=J_3^2 \quad , 
\qquad {\cal P}_{{\bf k},2}^\pm=1-J_3^2$}\vline \\ \hline\hline
  & longitudinal & mixed & transverse \\ \hline\hline
$\lambda_{k,r}$ $(n_r)$ & \;\; $\cos^2\theta$ (8), 0 (4) \;\;   & 1/2 (8), 0 (4)& 
$\sin^2\theta$ (8), 0 (4)
\\ \hline
$a_r$ & 1, 0 & 1, 0 & 1, 0 \\ \hline
$\overline{d}$& 6 & \;\; $5$ \;\; & 9/2\\ \hline
$\overline{\zeta}$ & -1/3& $-\ln\sqrt{2}$ & $\ln 2 - 5/6$ \\ \hline
\end{tabular}
\caption{Relevant quantities for the polar phase.}
\label{tablepolar}
\end{table}
 
\begin{table}  
\begin{tabular}[t]{|c||c|c|c|}
\hline
 & \multicolumn{3}{c}{{\bf Planar phase}}\vline \\ \hline\hline
 & \multicolumn{3}{c}{${\cal M}_{\bf k} = J_1[\a\,\hat{k}_1 + 
\b\,\g_{\perp,1}(\uk)]+ J_2[\a\,\hat{k}_2 + \b\,\g_{\perp,2}(\uk)]$}
\vline \\ \hline
 & \multicolumn{3}{c}{$L_{\bf k}^\pm = J_1^2 A_1 + J_2^2 A_2 + \{J_1,J_2\}\,B + 
[J_1,J_2]\,Z$}  \vline \\  \hline
 & \multicolumn{3}{c}{$\lambda_{k,1}=\a^2\sin^2\theta+\b^2(1 + \cos^2\theta)\qquad 
(n_1=8)\quad, \qquad \lambda_{k,2} = 0 \qquad (n_2=4)$}
\vline \\ \hline
 & \multicolumn{3}{c}{${\cal P}_{{\bf k},1}^\pm=L_{\bf k}^\pm/\lambda_{k,1} \quad , 
\qquad 
{\cal P}_{{\bf k},2}^\pm=1-L_{\bf k}^\pm/\lambda_{k,1}$}\vline \\ \hline\hline
  & longitudinal & mixed & transverse \\ \hline\hline
$\lambda_{k,r}$ $(n_r)$ & \;\; $\sin^2\theta$ (8), 0 (4)\;\;  & \;\; 1 (8),  0 (4)
\;\; & $1 + \cos^2\theta$ (8), 0 (4)
\\ \hline
$a_r$ & 1, 0 & 1, 0 & 1, 0 \\ \hline
$\overline{d}$& 6 &  21/4  & 9/2\\ \hline
$\overline{\zeta}$ & $\ln 2 - 5/6$ & 0 & $\ln\sqrt{2}-7/12+\pi/8$  \\ \hline
\end{tabular}
\caption{Relevant quantities for the planar phase.}
\label{tableplanar}
\end{table}
  
\begin{table}  
\begin{tabular}[t]{|c||c|c|c|}
\hline
 & \multicolumn{3}{c}{{\bf A phase}}\vline \\ \hline\hline
 & \multicolumn{3}{c}{${\cal M}_{\bf k} = J_3\,\{\a\,(\hat{k}_1+i\hat{k}_2)+
\b\,[\g_{\perp,1}(\uk)+i\g_{\perp,2}(\uk)]\}$}
\vline \\ \hline
 & \multicolumn{3}{c}{$L_{\bf k}^\pm = J_3^2[(A_1 + A_2)\pm 2i\,Z]$}  
\vline \\  \hline
 & \multicolumn{3}{c}{$\lambda_{k,1/2}=\a^2\sin^2\theta+\b^2(1 + \cos^2\theta) \pm
2\,\b\,\sqrt{\a^2\sin^2\theta+\b^2\cos^2\theta} \qquad 
(n_1=n_2=4)$ \,\, ,}
\vline \\ 
 & \multicolumn{3}{c}{$\lambda_{k,3} = 0  \qquad (n_3=4)$}\vline \\  \hline
 & \multicolumn{3}{c}{${\cal P}_{{\bf k},1}^\pm = \frac{1}{2} J_3^2 
(1\pm\frac{i}{\sqrt{A_1A_2-B^2}}\,Z) \,\, , \qquad  
{\cal P}_{{\bf k},2}^\pm = \frac{1}{2} J_3^2 (1\mp\frac{i}
{\sqrt{A_1A_2-B^2}}\,Z) \,\, , \qquad {\cal P}_{{\bf k},3}^\pm = 1 -J_3^2$}
\vline \\ \hline\hline
  & longitudinal & mixed & transverse \\ \hline\hline
$\lambda_{k,r}$ $(n_r)$ & \;\; $\sin^2\theta$ (8), 0 (4) \;\;  & \;\; 2 (4), 0 (8)
\;\; & $(1 + |\cos\theta|)^2$ (4), $(1 - |\cos\theta|)^2$ (4),  0 (4)
\\ \hline
$a_r$ & 1, 0 & 1, 0 & \;\; $1/2 + |\cos\theta|/(1+\cos^2\theta)$, 
$1/2 - |\cos\theta|/(1+\cos^2\theta)$, 0 \;\;\\ \hline
$\overline{d}$& 6 & 21/4 & 9/2\\ \hline
$\overline{\zeta}$ & $\ln 2 - 5/6$ & $\ln\sqrt{2}$ & $\ln2-1/3$  \\ \hline
\end{tabular}
\caption{Relevant quantities for the A phase.}
\label{tableA}
\end{table}

\begin{table}  
\begin{tabular}[t]{|c||c|c|c|}
\hline
 & \multicolumn{3}{c}{{\bf CSL phase}}\vline \\ \hline\hline
 & \multicolumn{3}{c}{${\cal M}_{\bf k} = {\bf J}\cdot[\a\,\uk+
\b\,\gperp(\uk)]$}
\vline \\ \hline
 & \multicolumn{3}{c}{$(L_{\bf k}^\pm)_{ab} = (\a^2+2\b^2)\,\d_{ab} 
- [\a\,\hat{k}_b+\b\,\g_{\perp,b}(\uk)]
[\a\,\hat{k}_a-\b\,\g_{\perp,a}(\uk)]$}  
\vline \\  \hline
 & \multicolumn{3}{c}{$\lambda_{k,1/2} = \frac{1}{2}\a^2+
2\b^2\pm\frac{1}{2}\a\sqrt{\a^2+8\b^2}\qquad 
(n_1=n_2=4) \quad, \qquad
\lambda_{k,3} = \a^2  \qquad (n_3=4)$}
\vline \\ \hline
 & \multicolumn{3}{c}{${\cal P}_{{\bf k},r}^\pm = \prod_{s\neq r}^3 
\frac{L_{\bf k}^\pm-\lambda_{k,s}}{\lambda_{k,r}-\lambda_{k,s}}$} \vline \\ \hline\hline
  & longitudinal & mixed & transverse \\ \hline\hline
$\lambda_{k,r}$ $(n_r)$ &\;\; 1 (8), 0 (4)\;\;  &\;\; 2 (4), 1/2 (8)\;\; &
 2 (8), 0 (4)\\ \hline
$a_r$ & 1, 0 & 2/3, 1/3 & 1, 0 \\ \hline
$\overline{d}$& 6 &  5  & 9/2\\ \hline
$\overline{\zeta}$ & 0 & $\ln\sqrt{2^{1/3}}$ & $\ln\sqrt{2}$  \\ \hline
\end{tabular}
\caption{Relevant quantities for the CSL phase.}
\label{tableCSL}
\end{table}

\begin{figure}[ht] 
\begin{center}
\includegraphics[width=13cm]{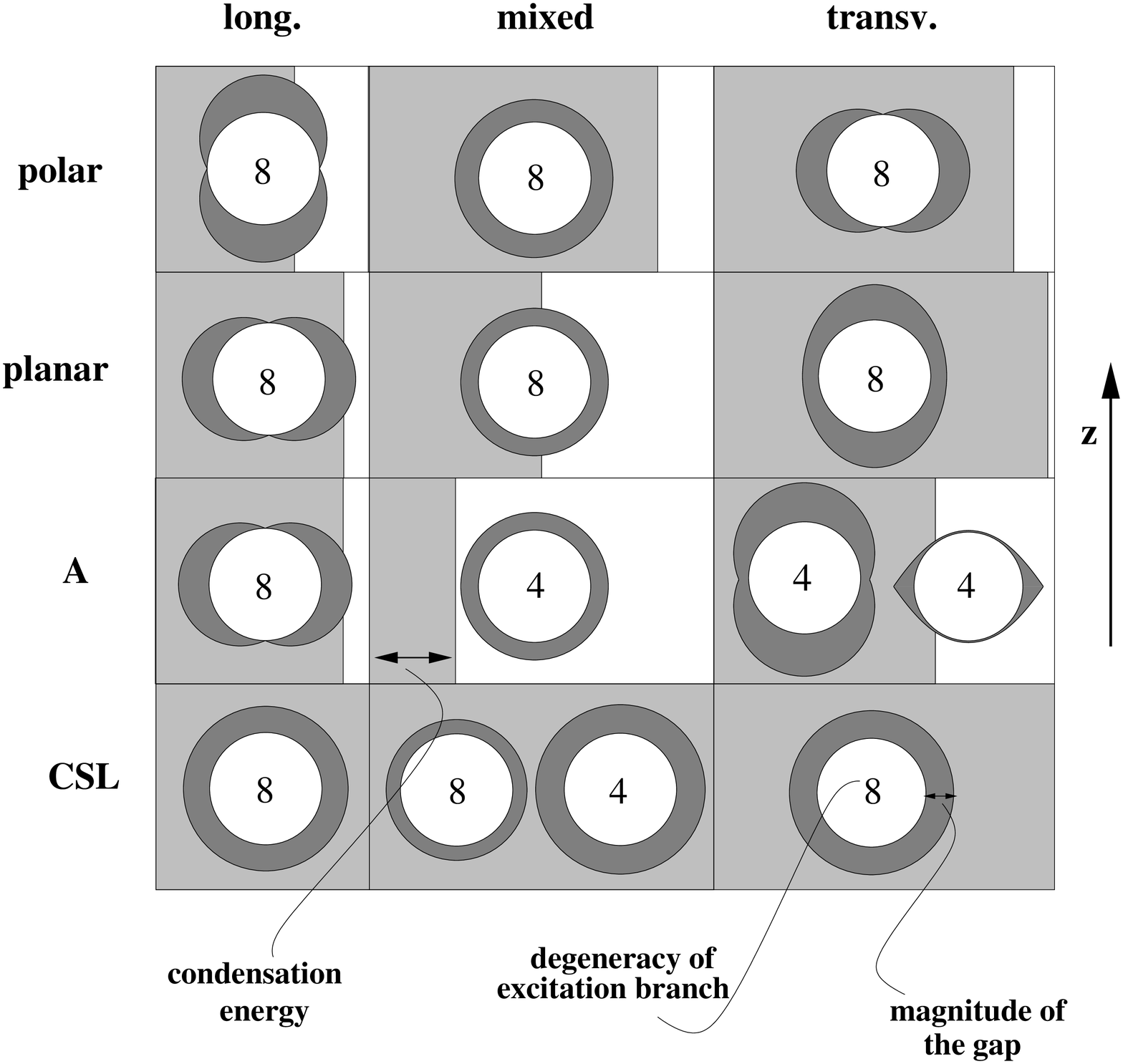}
\vspace{0.5cm}
\caption[Gap functions in a spin-one color superconductor]
{Schematic representation of the gap functions 
$\sqrt{\lambda_{r,k}}\,\phi_0$, given in Table \ref{tablegaps}, and  
condensation energies, given in Table \ref{tablepressure}. 
All gap functions are symmetric with respect to 
rotations around the $z$-axis. The numbers correspond to the degeneracies
of the respective excitation branches. Only the gapped excitations are shown. 
The condensation energies are drawn to scale in each column separately. }
\label{figuregaps}
\end{center}
\end{figure}

We summarize the relevant quantities for the different 
phases in Tables \ref{tablepolar} -- \ref{tableCSL}. The physically
important results are collected in Tables \ref{tablegaps} -- 
\ref{tablepressure} and Fig.\ \ref{figuregaps}. 

Let us first comment on Tables \ref{tablepolar} -- \ref{tableCSL}. For the
planar and A phases, we use the abbreviations
\begin{subequations} \label{defs}
\bea
A_{1/2} &\equiv& (\a^2-\b^2)\,\hat{k}_{1/2}^2 + \b^2 \,\, , \label{A12}\\
B&\equiv&(\a^2-\b^2)\,\hat{k}_1\hat{k}_2 \,\, ,\\
Z&\equiv&\b\left\{\a\left[\hat{k}_2\,\g_{\perp,1}(\uk)-  
\hat{k}_1\,\g_{\perp,2}(\uk)\right] - \b\left[\g_{\perp,1}(\uk)\,\g_{\perp,2}(\uk)-
\hat{k}_1\hat{k}_2\right]\right\} \,\, .
\eea
\end{subequations}
The quantities $A_{1/2}$, $B$, and $Z$ are diagonal in color space. But
while $A_{1/2}$ and $B$ are scalars, $Z$ is a nontrivial $4\times 4$ matrix 
in Dirac space. One can verify the relation
\be \label{Zsquared}
Z^2=B^2 -A_1A_2 \,\, .
\ee
Moreover, in the table for the planar phase, we denote the anticommutator 
by $\{-,-\}$. The angle $\theta$, used in Tables \ref{tablepolar} -- 
\ref{tableA} is the angle between the quark momentum 
${\bf k}$ and the $z$-axis. The indices $a,b \le 3$ in the second line of
Table \ref{tableCSL} are color indices.

The specific form of the matrices ${\cal M}_{\bf k}$ is obtained by inserting
the respective order parameters $\Delta$, cf.\ Fig.\ \ref{symmetries}, 
into the definition (\ref{Mk}). The matrices $L_{\bf k}^\pm$ 
are given by the definition (\ref{defL}). The calculation of their eigenvalues
$\lambda_{k,r}$ and corresponding degeneracies $n_r$ is presented in Appendix
\ref{AppA} for the planar and A phases. The same method can be applied to the
other phases, see for instance Refs.\ \cite{schmitt,thesis}. After 
determining the eigenvalues, 
the corresponding projectors are obtained with the help of Eq.\ (\ref{defP}).
All these quantities, ${\cal M}_{\bf k}$, $L_{\bf k}^\pm$, $\lambda_{k,r}$, $n_r$,
and ${\cal P}_{{\bf k},r}^\pm$, are computed for arbitrary linear combinations
of longitudinal and transverse gaps and are given in the first four lines 
of the Tables \ref{tablepolar} -- \ref{tableCSL}. The last four lines of these 
tables are devoted to the special cases of pure longitudinal and transverse 
gaps and the mixed gap. First, we present the eigenvalues $\lambda_{k,r}$ with 
the corresponding degeneracies, immediately deduced from the general ones. 
Then, we compute the quantities $a_r$, see Eqs.\ (\ref{eta}), and 
$\overline{d}$,
see Eqs.\ (\ref{d}) and (\ref{zeta}). The calculation of these quantities 
is more or less
straightforward, but may turn out to be lengthy. It involves performing
the color and Dirac traces in the quantities 
${\cal T}_{00}^s({\bf k},{\bf q})$ and ${\cal T}_t^s({\bf k},{\bf q})$ as well
as the angular integration over $\uk$, see Eq.\ (\ref{pullout}). For 
the most complicated cases, the transverse polar, planar, and A phases, we present
details of the calculation in Appendix \ref{transversephases}. 
Furthermore, in Appendix \ref{AppB} it is shown that $\overline{d}=d=6$ 
is a universal result for all longitudinal phases. A special
case is the mixed polar phase, which is the only case where $d$ turns out to 
be angular-dependent, $d=3(3+\cos^2\theta)/2$. This result has already been
obtained in Refs.\ \cite{schmitt,thesis}. However, it has not been realized 
that $\overline{d}=5$, and not $d$, enters the value of the gap $\phi_0$. 
Finally, the constant $\overline{\zeta}$ is straightforwardly obtained by 
using Eqs.\ (\ref{zetaiso}) and (\ref{zeta}).

\begin{table} 
\begin{center}
\begin{tabular}[t]{|c||r|c|r|}
\hline 
$\sqrt{\lambda_{k,r}}\phi_0/\phi_0^{\rm 2SC}$ & longitudinal $\quad$ & mixed & 
transverse 
$\quad\qquad\,\,\,\;\;\;$   \\ \hline\hline
polar & $|\cos\theta|\,e^{1/3}\, e^{-6}\;\;$ & $\;\;e^{-5}\;\;$ 
& 
$|\sin\theta|\,\frac{1}{2}\,e^{5/6}\,e^{-9/2}\;\;$ \\ \hline
planar & $\;\;|\sin\theta|\,\frac{1}{2}\,e^{5/6}\, e^{-6}\;\;$ & $e^{-21/4}$ & 
$\;\;\sqrt{1+\cos^2\theta}\,\frac{1}{\sqrt{2}}\,e^{7/12-\pi/8}
\,e^{-9/2}\;\;$ \\
\hline
A & $|\sin\theta| \, \frac{1}{2}\,e^{5/6}\,e^{-6}\;\;$ & $e^{-21/4}$ & 
$(1\pm|\cos\theta|)\,\frac{1}{2}\,e^{1/3}\,e^{-9/2}\;\;$ \\
\hline
CSL & $e^{-6}\;\;$ & $\;\;2^{(-1\pm 3)/6}\, e^{-5}\;\;$ & $e^{-9/2}\;\;$ \\ \hline

\end{tabular}
\caption[Gap functions]{Gap functions $\sqrt{\lambda_{k,r}}\phi_0$ in units
of the 2SC gap. Fig.\ \ref{figuregaps} illustrates this table schematically.}
\label{tablegaps} 
\end{center}
\end{table}

\begin{table}
\begin{center}
\begin{tabular}[t]{|c||c|c|c|}
\hline 
$T_c/T_c^{\rm 2SC}$ & $\;\;$ longitudinal $\;\;$ & mixed & $\;\;$ transverse  
$\;\;$ \\ \hline\hline
polar & $e^{-6}$ & $\;\;e^{-5}\;\;$ & 
$e^{-9/2}$ \\ \hline
planar & $e^{-6}$ & $\;\;e^{-21/4}\;\;$ & $e^{-9/2}$ \\
\hline
A & $e^{-6}$ & $e^{-21/4}$ & $e^{-9/2}$ \\
\hline
CSL & $e^{-6}$ & $e^{-5}$ & $e^{-9/2}$ \\ \hline

\end{tabular}
\caption{The critical temperature $T_c$ in units of the critical temperature
for the 2SC phase.}
\label{tableTc}
\end{center}
\end{table}

\begin{table}  
\begin{center}
\begin{tabular}[t]{|c||r|c|r|}
\hline
$\Delta p/\Delta p_{\rm 2SC}$ & longitudinal$\qquad\;\;$ & mixed & transverse
$\qquad\quad\,\;$ 
\\ \hline\hline
polar & $\,\,\frac{1}{3}\,e^{2/3}\,e^{-12}\simeq 0.65 \,e^{-12} \;\;$ & 
$\;\;e^{-10}\;\;$ & 
$\frac{1}{6}\,e^{5/3}\,e^{-9}\simeq 0.88\,e^{-9}\;\;$ \\ \hline
planar & $\;\;\frac{1}{6}\,e^{5/3}\,e^{-12}\simeq 0.88\,e^{-12}\;\;$ & 
$e^{-21/2}$ & $\;\;\frac{2}{3}\,
e^{7/6-\pi/4}\,e^{-9}\simeq 0.98\,e^{-9}\;\;$ \\ \hline
A & $\frac{1}{6}\,e^{5/3}\,e^{-12}\simeq 0.88\,e^{-12}\;\;$ & 
$\,\,\frac{1}{2} \, e^{-21/2}\,\,$ & $\,\,
\frac{1}{3}\,e^{2/3}\, e^{-9}\simeq 0.65\,e^{-9}\;\;$ 
\\ \hline
CSL & $e^{-12}\;\;$ & $\;\;3\cdot 2^{-4/3} \, e^{-10}\;\;$ & $e^{-9}\;\;$ 
\\ \hline
\end{tabular}
\caption[Zero-temperature pressure]{Zero-temperature pressure $\Delta p$ 
(= condensation energy) in units of the 2SC pressure. The results
are illustrated in Fig.\ \ref{figuregaps}.}
\label{tablepressure}
\end{center}
\end{table}

Let us now turn to the physically important results, Tables \ref{tablegaps}
-- \ref{tablepressure} and Fig.\ \ref{figuregaps}. In Table \ref{tablegaps}
we present the (angular-dependent) gap functions for zero temperature at the 
Fermi surface, as they occur in 
the quasiparticle energies, Eq.\ (\ref{excite}). They involve the square root
of the eigenvalue $\lambda_{k,r}$ and the factors $e^{-\overline{d}}$ and 
$e^{-\overline{\zeta}}$,
as shown in the solution of the gap equation, cf.\ Eq.\ (\ref{gapgeneral}).
The results are easily computed using the results of the previous tables.   
The magnitude of the gaps are reduced compared to the gaps in the spin-zero
phases by factors of the order $e^{-6}\simeq 2.4\cdot 10^{-3}$ through
$e^{-9/2}\simeq 1.1\cdot 10^{-2}$. Consequently, assuming the spin-zero
gaps to be of the order of 10\,MeV, the spin-one gaps are of the order 
of 10 -- 100\,keV. 

The angular structure of the gap functions, given in Table \ref{tablegaps},
is illustrated in Fig.\ \ref{figuregaps}.
In this figure, the magnitude of the gap at the Fermi surface is shown
as a function of the polar angle $\theta$. None of the functions depends
on the azimuthal angle $\varphi$, i.e., all figures are symmetric with 
respect to rotations around the $z$-axis. First note that the gaps in 
the CSL phase are isotropic. Nevertheless, there is a hidden anisotropy also 
in this phase, since the residual symmetry is not the group of rotations 
in real space but the group of joint rotations in color and real space.
Thus, analogous to the B phase in $^3$He one might call this phase 
``pseudoisotropic''. Other analogies to $^3$He can be found, particularly 
in the first column of the figure, representing the longitudinal gaps. 
All structures correspond to their analogues in $^3$He. This is plausible, 
because the respective gap matrices do not involve the nontrivial 
Dirac part $\gperp(\uk)$, and thus can, in this respect, be considered as 
the nonrelativistic limit.  Note that the longitudinal 
polar phase has 
a nodal line at the equator of the Fermi sphere, while the longitudinal 
planar and A phases have nodal points at the north and south pole of the
Fermi sphere. All longitudinal phases have one gapped 
and one ungapped excitation branch with degeneracies eight and four, 
respectively, cf.\ Appendix \ref{AppB}. Note that the degeneracies add up to 
12, since the matrix
$L_{\bf k}^+$ is a $12\times 12$ matrix, involving antiparticle degrees of 
freedom. Nevertheless, the physical 
degeneracies of the gapped branches have to be reduced by a factor 2 
compared to Fig.\ \ref{figuregaps}, since the antiparticle gaps are 
negligibly small.    

All phases shown in the second and 
third columns of the figure, have no analogues in $^3$He, because 
$\gperp(\uk)$ gives rise to a nontrivial Dirac structure.
The mixed gaps all are isotropic. The transverse gaps, however, exhibit 
nontrivial angular structures. The transverse polar phase has point nodes,
similar to the longitudinal planar and A phases. The transverse planar phase
has an anisotropic gap, however, the gap vanishes nowhere. The transverse 
A phase has one ungapped excitation and two different gapped ones, each with 
a nontrivial angular structure. One of these structures has no nodes but 
minima at the equator of the Fermi sphere, while the other has point nodes at 
the north and the south pole. 

The critical temperatures, presented in Table \ref{tableTc}, are obtained 
from Eq.\ (\ref{nonbcs}). As remarked below that equation, the absolute
value of $T_c$ depends only on the factor $e^{-\overline{d}}$. Therefore, the 
results of Table \ref{tableTc} are easy to interpret. The constant 
$\overline{d}$ assumes the value $\overline{d}=6$ for all longitudinal gaps, 
which therefore all have 
the same transition temperature, being of the order of 10\,keV 
(again applying $\phi_0^{\rm 2SC}\simeq 10$ MeV). Since
the mixed phases have different values for $\overline{d}$, their transition 
temperatures
differ, ranging in the order 30\,keV -- 40\,keV. The transverse phases have 
the largest transition temperature, $T_c\simeq 60$\,keV.  
  
The most important physical results of this paper can be found in
Table \ref{tablepressure}, and, schematically, in Fig.\ \ref{figuregaps}, 
since they answer the question of the preferred
state in a spin-one color superconductor at zero temperature. 
Crucial to this question is
the factor $e^{-\overline{\zeta}}$ in the gap function, which is strongly 
affected by the nontrivial angular structures in the gap equation, which was
not realized in Ref.\ \cite{thesis}, where a simple approximation for 
the angular integral had been used. The first obvious result, already 
indicated by the results of Ref.\ \cite{schaefer} and independent of 
the constant $\overline{\zeta}$, is that the pressure of all transverse 
phases is larger
than that of the mixed phases, which, in turn, is larger than that of the
longitudinal phases. Therefore, the most interesting and relevant 
phases are the transverse ones. Nevertheless, let us mention, that in the 
results
for the longitudinal phases we recover Eq.\ (\ref{ratioAB}), i.e., the
ratio between the condensation energies of the longitudinal A and CSL phases 
is identical to the corresponding one of the A and B phases in $^3$He.
For the transverse phases, the result cannot be anticipated from the
theory of $^3$He. We find that the condensation energy of the phases with
nodal lines or points of the gap, i.e., the polar and A phases, is smaller than that
of the phases without nodes. However, note that the polar phase has a larger 
condensation energy than the A phase, although it has more excitation 
branches with nodes in the gap function (eight compared to four).
The two phases without zeros in the gap function, i.e., the planar and CSL phases,
differ only by two percent in their condensation energy. The isotropic transverse 
CSL phase has the largest condensation energy.   

Note that electric charge neutrality of the system does not affect the
results, since we are considering a one-flavor system. This, of course,
is in contrast to the systems where quarks of different flavors form Cooper
pairs. In these systems, for instance in the (gapless) 2SC and CFL phases,
charge neutrality plays an important role regarding the question of the 
ground state. The effect of the condition of {\em color} charge neutrality 
for a one-flavor system is less trivial. In the polar and A phases, Cooper 
pairs carry color charge anti-blue, i.e., only red and green quarks condense while 
the blue quarks remain unpaired. This pattern is known from the 2SC phase.
It leads to a nonvanishing expectation 
value of the gluon field $A_8^\mu$ \cite{neutrality}, which ensures color 
neutrality of the system. This is equivalent 
to introducing a color chemical potential $\mu_8$ \cite{shovkovy} which 
discriminates between the chemical potentials for paired and 
unpaired quarks. In the planar phase, Cooper pairs are formed by red/blue and 
green/blue quarks, i.e., they carry color charge anti-green or anti-red. 
Also in this case, one expects the color chemical potential for red and green quarks 
to differ from that for the blue quarks because of the residual color group $SU(2)$. 
Consequently, quarks of different chemical 
potentials form Cooper pairs. In both cases, the condition of color neutrality
might reduce the pressure. In contrast, the CSL phase automatically fulfills color 
neutrality, since there are equal numbers of Cooper pairs with 
colors anti-red, anti-green, and anti-blue. Therefore, the result 
that the transverse CSL phase has the largest pressure 
is insensitive with respect to electric and color neutrality conditions.

\section{Summary and outlook} \label{conclusions}

In this paper, we have investigated the possible color-superconducting phases 
in cold and dense quark matter, where quarks of the same flavor form Cooper
pairs, which leads to Cooper pairs with total spin one. It has been argued 
that these phases might be relevant in the interior of neutron stars, since a 
mismatch of Fermi surfaces, if too large, forbids pairing of quarks of 
different flavors. Since the structure of the order parameter of the 
spin-one phases corresponds to a complex $3 \times 3$ matrix, a priori 
a multitude of phases seems to be possible. It has been the main goal of this
paper to pick the order parameter that leads to the phase with the largest 
pressure, indicating that it is the favored spin-one color superconductor.
In principle, this task is comparable to the three-flavor case, where the
CFL phase turns out to be the favored one. However, in that case, the 
argument that favors the CFL phase is more or less obvious and plausible, since
it is the only possible phase in which all quasiparticles attain a gap 
in their excitation spectrum. The present paper shows that for the spin-one
phases the argument is more subtle, since, in particular, it involves
the angular structures of the gap functions. Angular-dependent gap functions
are known from condensed-matter systems, for instance from superfluid 
$^3$He. We have shown that a certain class of spin-one phases, named 
longitudinal phases (where quarks of the same chirality form Cooper pairs),
reproduces some features of the phases in $^3$He, namely the angular 
structure and the relative condensation energies of the several phases.
However, most of the investigated phases, and in particular the preferred 
transverse phases (where quarks of opposite chirality form Cooper pairs),
have no analogues in $^3$He. 

The specific technical and physical results of the paper can be summarized as 
follows. In the systematic classification of order parameters, based on
group-theoretical arguments, we have found four phases with uniquely 
defined order parameter matrices. These phases, the polar, planar, A, and CSL 
phases, all break color and rotational symmetries. From the 
theory of $^3$He we transfer the conclusion that these order 
parameters correspond to the inert phases, although the 
different group structure in principle demands for a more
careful investigation of this statement. Nevertheless, we have picked the 
above mentioned four phases for a more detailed discussion. Before 
evaluating these phases specifically, we have presented a general solution of 
the QCD gap equation. This treatment generalizes the one presented in 
Ref.\ \cite{schmitt}. The main conclusion is that an angular dependence
of the gap function gives rise to a nontrivial factor $e^{-\overline{\zeta}}$ 
in the expression for the zero-temperature gap. The explicit result
of the gap is
\be 
\phi_0 = 2\,\tilde{b}\,b_0'\,e^{-\overline{d}}\,e^{-\overline{\zeta}} \,\mu \, 
\exp\left(-\frac{\pi}{2\bar{g}}\right) \,\, .
\ee
It generalizes Eq.\ (\ref{gapconstant}) by the replacements
\be 
\zeta\to \overline{\zeta}\equiv 
\frac{\left\langle\ell_q\,\zeta\right\rangle_{\uq}}
{\left\langle \ell_q\right\rangle_{\uq}} 
\,\, , \qquad d\to \overline{d}\equiv 
\frac{\left\langle\ell_q\,d\right\rangle_{\uq}}
{\left\langle \ell_q\right\rangle_{\uq}} \,\, .
\ee 
The quantity $\ell_q$ is (half of) the sum of the 
eigenvalues $\lambda_{k,r}$ of the matrix $L_{\bf k}^+$, cf.\ Eq.\ (\ref{ell}). 
These 
eigenvalues contain the angular structure of the gaps. According to 
Eq.\ (\ref{zetafromL}), $\overline{\zeta}$ can be determined
solely from the quasiparticle spectrum.  
The physical meaning of the replacement $\zeta\to\overline{\zeta}$ 
can be illustrated as follows. Consider an isotropic gap function, $\phi$, 
and an anisotropic one,
say $|\sin\theta|\,\phi$, where $\theta$ is the angle between the quark 
momentum and one fixed spatial axis. One would immediately conclude that the 
condensation 
energy of the former one is larger, since the latter assumes its maximum 
value $\phi$ only for $\theta=\pi/2$, while it is reduced for all other 
angles. However, we have found that the latter one actually is equipped with 
the above mentioned factor, i.e., it reads 
$|\sin\theta|\,e^{-\overline{\zeta}}\,\phi$.
If $e^{-\overline{\zeta}}>1$, the question of the preferred phase is 
nontrivial and 
depends on the special value of $\overline{\zeta}$. The general result 
presented here reproduces the special cases of one or two isotropic gaps.  

Furthermore, we have computed a general expression for the critical 
temperature,
\be 
\frac{T_c}{\phi_0} = \frac{e^\g}{\pi}\; e^{\overline{\zeta}} \,\, .
\ee
This equation shows that the BCS relation $T_c/\phi_0\simeq 0.57$ is not only 
violated in the cases of a two-gap structure and a gapless 
color superconductor.
It is also violated in the case of a single gapped excitation branch if 
the corresponding gap is anisotropic. In this case, the ratio between 
the critical temperature and the quadratic mean $\overline{\phi}$ of the 
gap is given by Eq.\ (\ref{ratio}), which is exactly the same result as for
$^3$He. 

Finally, a general expression of the pressure at zero temperature has been 
derived starting from the effective action, which can be obtained from the 
QCD partition function using the CJT formalism. In particular, the pressure 
contains the condensation energy (density), which is the difference of 
the pressure in the superconducting phase compared to the normal-conducting 
phase,
\be 
\Delta p = \frac{\m^2}{16\,\pi^2}\,\sum_r n_r
\overline{\phi}_r^2 \,\, ,
\ee
Physically, this result is plausible. It is equivalent to counting the 
gapped branches and weight them with the angular average of the square of the 
 corresponding gaps. Again, our 
general result reproduces the well-known results for the spin-zero gaps.

These general results for the gap, the critical temperature, and the pressure
have been applied to the above mentioned four spin-one phases. The genuinely
new results concern all phases with nontrivial angular structure. These are 
the longitudinal and transverse polar, planar, and A phases, while the 
``pseudoisotropic'' CSL phase already has been discussed in Refs.\ 
\cite{schaefer,schmitt}. It has been shown that several of the phases exhibit
nodal points or lines of the gap, similar to $^3$He and 
several high-$T_c$ superconductors in condensed-matter physics \cite{tsuei}.
The magnitude of the gap is dominated by the factor $e^{-\overline{d}}$ 
which varies
from $e^{-6}$ for longitudinal gaps to $e^{-9/2}$ for transverse gaps. These
factors have already been found in Refs.\ \cite{schaefer,ren}. The 
consequence is that the spin-one gaps are smaller by \mbox{2 -- 3} orders of 
magnitude
than the spin-zero gaps, which renders them of the order of 10 -- 100\,keV.
These numbers are not changed essentially by the factor 
$e^{-\overline{\zeta}}$ which,
in all cases, is of order one. Nevertheless, this factor is decisive for 
the question of the favored phase, which can be found among the 
transverse phases. Since the pressure 
of all transverse phases
contains the factor $e^{-9}$ and since the number of the gapped 
excitation branches in all transverse phases equals 8, the preferred phase 
is determined by the specific factors originating from the angular structure. 
The result is that the transverse CSL phase has the largest pressure. 
This phase has an isotropic energy gap, contrary to all other transverse phases.
It has been discussed that this result remains unchanged under the condition of overall
color charge neutrality. 

Let us finally add some remarks regarding astrophysical consequences. 
As a first result, we note that the existence of a spin-one color 
superconductor in the interior of a neutron star is not ruled out by
the temperatures, which, in the case of old neutron stars, is in the range 
of keV. In order
to find consequences of a spin-one color-superconducting core for the
observables of the neutron star, it seems to be very promising to study
the interplay of the superconductor with the magnetic field of the star.
In Refs.\ \cite{schmitt2,schmitt3} it has been discussed in detail that
the mixed polar as well as the mixed CSL phase exhibit an electromagnetic
Meissner effect, contrary to the spin-zero phases. From these results one
concludes that it is very likely that also the other spin-one phases
expel electromagnetic fields, although an explicit calculation has not yet been 
done. For the transverse CSL phase, this statement is obvious due to its symmetry
breaking pattern. 
It might be interesting to investigate whether the angular structure, in 
particular the nodal points or lines, leads to anisotropies regarding
external magnetic fields, for instance to angular-dependent penetration 
depths. Furthermore, external magnetic fields should be included into 
the discussion of the favored phase. As we know from $^3$He, above a certain
threshold for the external magnetic field, the B phase is no longer the ground 
state of the system. Instead, the anisotropic A phase becomes favored.
Therefore, it remains a physically important project for the future to
investigate if the CSL phase is still preferred in the presence of a magnetic field. 

Another measurable property of a neutron star which is likely to be affected
by a color-superconducting core is its temperature, or, more precisely, its
cooling curve (= temperature as a function of time) \cite{hotwater}. Recently, it has 
been argued, that, unlike a spin-zero color superconductor, a spin-one 
color superconductor could explain the observed cooling curves \cite{blaschke}.
The main ingredient of this conclusion has been the magnitude of the gap, i.e.,
the keV-gap compared to the MeV-gap of the 2SC or CFL phases. 
The physical mechanism behind this cooling curve is the emission of 
neutrinos, which dominates the cooling of the star after the first few 
seconds after its creation. Since one of the involved processes of neutrino 
emission is 
the Urca process, which requires the breaking of a Cooper pair, it might be 
interesting to take into account not only the effect of the magnitude but also 
of the nodal structure of the gap. Moreover, the nodes of the
gap function would certainly affect the specific heat of the system. 
Note for instance that in the A phase of $^3$He (which has the same 
nodal structure as the longitudinal A phase in a spin-one color 
superconductor), the specific heat depends on temperature according to a power
law, 
while in the B phase the temperature dependence of the specific heat shows
an exponential behavior. Similar effects can be expected for a 
spin-one color superconductor and have to be included into a careful
discussion of the cooling behavior of a neutron star.

\section*{Acknowledgments}

I thank D.H. Rischke for valuable suggestions and a careful reading
of the manuscript. I thank T.\ Sch\"afer, I.A.\ Shovkovy, and 
Q.\ Wang for valuable discussions and comments.

\begin{appendix}

\section{Order parameters corresponding to residual groups $H=U(1)\times H'$}
\label{orderparameters}

In this appendix, we evaluate the invariance equation (\ref{invarianceapp}).
This means that we determine all possible residual subgroups of the 
form $H=U(1)\times H'$ and the corresponding order parameters, using 
Eq.\ (\ref{generator1}) as an ansatz for the generator for the residual $U(1)$.
The invariance condition can be written as a system of nine equations,
\begin{subequations} \label{setofequations}
\bea
\left(-\frac{a_8}{2\sqrt{3}} + 2c\right)\D_{11} + ib_3\D_{12} &=& 0\,\, , \\
\left(-\frac{a_8}{2\sqrt{3}} + 2c\right)\D_{12} - ib_3\D_{11} &=& 0\,\, , \\
\left(-\frac{a_8}{2\sqrt{3}} + 2c\right)\D_{13} &=& 0 \,\, , \\
\left(-\frac{a_8}{2\sqrt{3}} + 2c\right)\D_{21} + ib_3\D_{22} &=& 0\,\, , \\
\left(-\frac{a_8}{2\sqrt{3}} + 2c\right)\D_{22} - ib_3\D_{21} &=& 0\,\, , \\
\left(-\frac{a_8}{2\sqrt{3}} + 2c\right)\D_{23} &=& 0 \,\, , \\
\left(-\frac{a_8}{\sqrt{3}} + 2c\right)\D_{31} + ib_3\D_{32} &=& 0\,\, , \\
\left(-\frac{a_8}{\sqrt{3}} + 2c\right)\D_{32} - ib_3\D_{31} &=& 0\,\, , \\
\left(-\frac{a_8}{\sqrt{3}} + 2c\right)\D_{33} &=& 0 \,\, .
\eea
\end{subequations}
The corresponding coefficient 
matrix $A$ exhibits a block structure and the determinant thus factorizes
into four sub-determinants. Therefore, we have to consider the equation 
\be 
0 = {\rm det} A = {\rm det} A_1\,{\rm det} A_2\,{\rm det} A_3\,
{\rm det}A_4 \,\, ,
\ee  
where
\begin{subequations}
\bea
{\rm det}A_1 &=& \left[\left(-\frac{a_8}{2\sqrt{3}} + 
2c\right)^2-b_3^2\right]^2    \,\, , \label{det1}\\
{\rm det}A_2 &=& \left(-\frac{a_8}{2\sqrt{3}}+2c\right)^2 \,\, ,\label{det2}\\
{\rm det}A_3 &=& \left(\frac{a_8}{\sqrt{3}}+2c\right)^2 - b_3^2 \,\, ,
\label{det3}\\
{\rm det}A_4 &=& \frac{a_8}{\sqrt{3}}+2c \,\, .\label{det4}
\eea
\end{subequations}
Now, one can systematically list all possibilities that yield a zero 
determinant of the coefficient matrix and thus allow for a nonzero order 
parameter.

\begin{enumerate}
\item ${\rm det}A_1 = 0$.

Here we distinguish between the cases (i) where the two terms in the angular 
brackets of Eq.\ (\ref{det1}) cancel each other and (ii) where they 
separately vanish.

\begin{enumerate}

\item $a_8$, $c$ arbitrary, $b_3 = -a_8/(2\sqrt{3}) + 2c$.

Inserting these conditions for the coefficients into Eqs.\ 
(\ref{setofequations}), one obtains for the order parameter matrix
\be
\D= \frac{1}{N}\left(\begin{array}{ccc} \D_1&i\D_1&0\\ \D_2&i\D_2&0\\0&0&0 
\end{array}\right) \,\, , \label{order1i} 
\ee
where the factor $1/N$ with  
$N=(2|\D_1|^2+2|\D_2|^2)^{1/2}$ accounts for the normalization 
(\ref{normalize2}). In this case, the order parameter contains 
two independent parameters $\D_1$ and $\D_2$. 
From this form of the order parameter, we can now determine the group $H'$
in Eq.\ (\ref{H'}). 
Inserting $\D$ into Eq.\ (\ref{invariance}) and using the fact that the 
parameters $\D_1$, $\D_2$ are independent of each other, one obtains 
the conditions
\be \label{conditions1i}
a_1 = \ldots = a_7 = b_1 = b_2 = 0 \,\, ,\quad 
\frac{1}{2\sqrt{3}}a_8 + b_3 -2c = 0 \,\,.
\ee
Consequently, 
\be \label{H1i}
H = U(1)\times U(1) \,\, ,
\ee 
since a vanishing coefficient in Eq.\ (\ref{conditions1i}) translates to 
a ``broken dimension''
of $G$. For instance, $a_1=0$ means that $T_1$ does not occur in the generators
of $H$, etc. The dimensions of the residual Lie group can be counted with
the help of the number of the conditions for the coefficients. 
Since ${\rm dim}\,G={\rm dim}\,G_1 + {\rm dim}\,G_2 + {\rm dim}\,G_3 
= 8+3+1=12$, 
and the number of conditions in Eqs.\ (\ref{conditions1i}) is 10, we
conclude ${\rm dim}\,H=2$, which is in agreement with Eq.\ (\ref{H1i}).
Or, in other words, there is an additional $U(1)$, i.e., $H'=U(1)$ in 
Eq.\ (\ref{H'}) because the equation relating
the three coefficients $a_8$, $b_3$, $c$ allows for two linearly independent 
generators $U$ and $V$ which are linear combinations of the 
generators $T_8$, $J_3$, ${\bf 1}$. Note that $U$ and $V$ are not uniquely 
determined. One possible choice is
\be \label{gen1i}
U = T_8 - \frac{1}{2\sqrt{3}}J_3 \,\, , \qquad V = 2J_3 + {\bf 1} \,\, .
\ee

Different order parameters are obtained from two subcases: 

First, 
one can impose the additional relation $c=-a_8/(2\sqrt{3})$ between the
two coefficients that have been arbitrary above. Then,
$b_3=-a_8\sqrt{3}/2$. These two conditions yield
\be
\D= \frac{1}{N}\left(\begin{array}{ccc} \D_1&i\D_1&0\\ \D_2&i\D_2&0\\0&0&\D_3 
\end{array}\right) \,\, , \label{order1iadd} 
\ee
where $N=(2|\D_1|^2+2|\D_2|^2+|\D_3|^2)^{1/2}$. In this case, 
Eq.\ (\ref{invariance}) leads to 11 conditions for the coefficients 
$a_m$, $b_n$, $c$, which leaves a subgroup
\be
H=U(1) \,\, ,
\ee
generated by a linear combination of generators of all three original 
subgroups $G_1$, $G_2$, $G_3$, 
\be
U= T_8 - \frac{\sqrt{3}}{2}J_3 - \frac{1}{2\sqrt{3}}{\bf 1} \,\, .
\ee

Second, one can set one of the coefficients $a_8$, $c$ to zero. 
The condition $c=0$ does not yield a new case. But $a_8=0$, and 
consequently $b_3=2c$, has to be treated separately. 
In this case, Eqs.\ (\ref{setofequations}) yield
\be
\D= \frac{1}{N}\left(\begin{array}{ccc} \D_1&i\D_1&0\\ \D_2&i\D_2&0\\ \D_3&
i\D_3 &0 
\end{array}\right) \,\, , \label{order1ii} 
\ee
where $N=(2|\D_1|^2+2|\D_2|^2+2|\D_3|^2)^{1/2}$. The residual group is given 
by
\be
H = U(1) \,\, ,
\ee
generated by
\be
U = 2J_3 + {\bf 1} \,\, .
\ee 

\item $c=a_8/(4\sqrt{3})$, $b_3=0$. 

Here, one obtains
\be
\D= \frac{1}{N}\left(\begin{array}{ccc} \D_1&\D_2&\D_3\\ \D_4&\D_5&\D_6\\ 
0& 0 &0  
\end{array}\right) \,\, , \label{order1iv} 
\ee
where $N=(\sum_{i=1}^6|\D_i|^2)^{1/2}$. Again, the residual group is
one-dimensional,
\be
H=U(1) \,\,  ,
\ee
generated by
\be
U = T_8 + \frac{1}{4\sqrt{3}}{\bf 1} \,\, .
\ee

\end{enumerate}

\item ${\rm det}A_2 = 0$.

This determinant vanishes in the following cases:

\begin{enumerate}
\item $b_3$ arbitrary, $c=a_8/(4\sqrt{3})$.

With Eqs.\ (\ref{setofequations}), one obtains
\be
\D= \frac{1}{N}\left(\begin{array}{ccc} 0&0&\D_1\\ 0&0&\D_2\\ 0&0&0 
\end{array}\right) \,\, , \label{order2i} 
\ee
where $N=(|\D_1|^2+|\D_2|^2)^{1/2}$. Inserting $\D$ into 
Eq.\ (\ref{invariance}) yields
\be \label{conditions2i}
a_1 = \ldots = a_7 = b_1 = b_2 = 0 \,\, ,\quad 
c=\frac{1}{4\sqrt{3}}a_8  \,\,.
\ee
As for the order parameter (\ref{order1i}), the residual group
is given by  
\be \label{H2i}
H = U(1)\times U(1) \,\, .
\ee 
However, the corresponding generators differ from those in Eqs.\ (\ref{gen1i}),
\be
U = T_8 + \frac{1}{4\sqrt{3}} {\bf 1} \,\, ,\qquad V=J_3 \,\, .
\ee 

\item $b_3$ arbitrary, $a_8=c=0$.

In this case,
\be 
\D= \frac{1}{N}\left(\begin{array}{ccc} 0&0&\D_1\\ 0&0&\D_2\\ 0&0&\D_3 
\end{array}\right) \,\, , \label{order2ii} 
\ee
where $N=(|\D_1|^2+|\D_2|^2+|\D_3|^2)^{1/2}$. The residual group is
\be 
H=U(1)\,\, ,
\ee
which is a subgroup of the spin group $G_2=SU(2)_J$, since it is
generated by 
\be
U=J_3 \,\, .
\ee
\end{enumerate}

\item ${\rm det}A_3 = 0$.
 
\begin{enumerate}
\item $a_8$, $c$ arbitrary, $b_3=a_8/\sqrt{3} + 2c$.

In this case, we find with Eqs.\ (\ref{setofequations}),
\be
\D= \frac{1}{\sqrt{2}}\left(\begin{array}{ccc} 0&0&0\\ 0&0&0\\ 1&i&0 
\end{array}\right) \,\, . \label{orderA} 
\ee
This matrix differs from all previously discussed order parameters in that 
it is uniquely determined. It corresponds to the A phase.
Here, as in all cases above, we omitted a possible phase factor which 
could multiply $\D$ without violating the normalization. 
Inserting $\D$ into Eqs.\ (\ref{invariance}) yields the following relations,
\be \label{conditions3i}
a_4 = \ldots = a_7 = b_1 = b_2 = 0 \,\, ,\quad 
\frac{1}{\sqrt{3}}a_8-b_3+2c=0 \,\,.
\ee
Consequently, besides the relation between $a_8$, $b_3$, and $c$, 
there are only 6
additional conditions. Thus, the dimension of the residual group is 
$12-7=5$. We obtain
\be
H = SU(2)\times U(1)\times U(1) \,\, ,
\ee
where $SU(2)$ is generated by $T_1$, $T_2$, $T_3$, and thus is a 
subgroup of the color gauge group $G_1=SU(3)_c$. 
For the generators of the two $U(1)$'s one can choose 
\be \label{gen3i}
U=T_8-\frac{1}{2\sqrt{3}}{\bf 1} \,\, , \qquad V=J_3+{\bf 1} \,\, .
\ee
As in case 1.(i), there is a subcase that produces an additional 
order parameter. Namely, if we require the condition $c=a_8/(4\sqrt{3})$, 
which yields $b_3=\sqrt{3}\,a_8/2$, we obtain
\be
\D= \frac{1}{N}\left(\begin{array}{ccc} 0&0&\D_2\\ 0&0&\D_3\\ \D_1&i\D_1&0 
\end{array}\right) \,\, , \label{order3i1} 
\ee
where $N=(2|\D_1|^2+|\D_2|^2+|\D_3|^2)^{1/2}$. From Eqs.\ (\ref{setofequations})
we conclude that all other coefficients vanish. Consequently,
\be
H=U(1) \,\, ,
\ee
with the generator
\be
U=T_8 + \frac{\sqrt{3}}{2}J_3 + \frac{1}{4\sqrt{3}}{\bf 1} \,\, .
\ee

\item $b_3 = 0$, $c=-a_8/(2\sqrt{3})$.

With Eqs.\ (\ref{setofequations}) one obtains
\be \label{order3ii}
\D=\frac{1}{N}\left(\begin{array}{ccc} 0&0&0\\0&0&0\\ \D_1&\D_2&\D_3 
\end{array}\right) \,\, ,  
\ee
where $N$ is defined as in Eq.\ (\ref{order2ii}). From Eq.\ (\ref{invariance})
we conclude in this case 
\be
a_4 = \ldots = a_7= b_1=b_2=b_3=0 \,\, , \quad c=-\frac{a_8}{2\sqrt{3}} \,\, .
\ee
Hence, the residual group is
\be
H = SU(2)\times U(1) \,\, ,
\ee
generated by $T_1$, $T_2$, $T_3$, and
\be
U = T_8 - \frac{1}{2\sqrt{3}}{\bf 1} \,\, .
\ee

\end{enumerate}

\item ${\rm det}A_4 = 0$.

There are two cases in which ${\rm det}A_4=0$:

\begin{enumerate}
\item $b_3$ arbitrary, $c=-a_8/(2\sqrt{3})$. 

These relations lead to 
\be
\D=\left(\begin{array}{ccc} 0&0&0\\0&0&0\\0&0&1 
\end{array}\right) \,\, , \label{orderpolar} 
\ee
As in the A phase, Eq.\ (\ref{orderA}), the order parameter is uniquely 
determined. It describes the polar phase. Inserting the order parameter 
of the polar phase 
into the invariance condition, Eq.\ (\ref{invariance}), yields
\be
a_4 = \ldots = a_7 = b_1 = b_2 = 0 \,\, ,\quad 
c=-\frac{1}{2\sqrt{3}}a_8 \,\,.
\ee
Therefore, the residual group is
\be 
H=SU(2)\times U(1)\times U(1)
\ee
with the generators $T_1$, $T_2$, $T_3$, and 
\be \label{genpolar}
U=T_8-\frac{1}{2\sqrt{3}}{\bf 1} \,\, , \qquad V=J_3 \,\, .
\ee
Thus, the symmetry breaking pattern in the polar phase is similar to that
in the A phase, cf.\ (\ref{gen3i}). But while in the A phase both 
residual $U(1)$'s are combinations of the original symmetries, 
in the polar phase, one of them is a subgroup of $G_2=SU(2)_J$.

\item $b_3$ arbitrary, $a_8=c=0$.

This case is identical to case 2.(ii).

\end{enumerate}
\end{enumerate}

\section{Eigenvalues of $L_{\bf k}^\pm$ in the planar and A phases} 
\label{AppA}

\subsection{Planar phase} \label{AppA1}

In order to compute the eigenvalues of the matrix 
$L_{\bf k}^+ = L_{\bf k}^-$, given in Table 
\ref{tableplanar}, we use Eq.\ (\ref{Zsquared}) and the (anti)commutation 
properties of
the color matrices $(J_i)_{jk}=-i\e_{ijk}$ to obtain 
$(L_{\bf k}^\pm)^2 = (A_1 + A_2)\,L_{\bf k}^\pm$. Therefore,
\be \label{npower}
(L_{\bf k}^\pm)^n = (A_1 + A_2)^{n-1}\,L_{\bf k}^\pm \,\, .
\ee
This simple relation for the $n$-th power of $L_{\bf k}^\pm$ 
can be used to find the roots of the equation 
\be
{\rm det}(\lambda - L_{\bf k}^\pm)=0 \,\, ,
\ee
which yield the eigenvalues $\lambda$ of $L_{\bf k}^\pm$. Since
\be
{\rm det} \left( \lambda - L_{\bf k}^\pm \right) \equiv \exp
\left[ {\rm Tr} \,  \ln \left( \lambda - L_{\bf k}^\pm  \right) 
\right] \,\, ,
\ee
we consider the trace of the logarithm of $\lambda - L_{\bf k}^\pm$,
\be
{\rm Tr}  
\ln \left( \lambda - L_{\bf k}^\pm  \right)  = \ln \lambda 
\, {\rm Tr}\, {\bf 1}
+ {\rm Tr}  \ln \left( {\bf 1} - \frac{L_{\bf k}^\pm}{\lambda} \right) 
= \ln \lambda \, {\rm Tr}\, {\bf 1} 
- \sum_{n=1}^{\infty} \frac{1}{n}\, \lambda^{-n} \,
{\rm Tr}\, (L_{\bf k}^\pm)^n \,\, .
\ee
The second term on the right-hand side can be evaluated using 
Eq.\ (\ref{npower}). With $\Tr L_{\bf k}^\pm = 8(A_1 + A_2)$ one obtains
\be
{\rm det}(\lambda - L_{\bf k}^\pm) = \lambda^4[\lambda-(A_1 + A_2)]^8 \,\, ,
\ee
which yields the eigenvalues $A_1 + A_2$ and $0$ with degeneracies 8 and 4, 
respectively. With the definition (\ref{A12}), this confirms the 
entries for $\lambda_{k,r}$ in Table \ref{tableplanar}.

\subsection{A phase} \label{AppA2}

In order to derive an expression for the $n$-th power of the matrix 
$L_{\bf k}^\pm$, given in Table \ref{tableA}, we note that
\be \label{Lsquared}
(L_{\bf k}^\pm)^2 = 2(A_1 + A_2)L_{\bf k}^\pm - [(A_1-A_2)^2+4B^2] J_3^2\,\, ,
\ee
where Eq.\ (\ref{Zsquared}) has been used. Since $A_{1/2}$ and $B$ are scalars
and $J_3^2 L_{\bf k}^\pm = L_{\bf k}^\pm$, we have
\be
(L_{\bf k}^\pm)^n = a_n L_{\bf k}^\pm +  b_n J_3^2 \,\, ,
\ee
with real coefficients $a_n$, $b_n$. Applying Eq.\ (\ref{Lsquared}),
one derives the following recursion relations for these coefficients,
\be
a_{n+1} = 2(A_1 + A_2)\,a_n + b_n \,\, , \qquad 
b_{n+1} = -[(A_1-A_2)^2 + 4B^2]\,a_n \,\, .
\ee
With the ansatz of a power series, $a_n = p^n$, one obtains a quadratic 
equation for $p$, which has the two solutions 
\be
p_{1/2} = A_1 + A_2 \pm 2\sqrt{A_1A_2 -B^2} \,\, .
\ee
Therefore, $a_n$ is a 
linear combination of the $n$-th powers of these solutions, 
$a_n=\eta_1 p_1^n + \eta_2 p_2^n$. The coefficients $\eta_1$, $\eta_2$ can be
determined from $a_1=1$, $a_2=2(A_1+A_2)$. One finds
\be 
a_n = \frac{1}{4}\frac{1}{\sqrt{A_1A_2-B^2}}(p_1^n - p_2^n) \,\, ,\qquad
b_n = -\frac{1}{4}\frac{(A_1 - A_2)^2 + 4B^2}{\sqrt{A_1A_2-B^2}}
(p_1^{n-1} - p_2^{n-1}) \,\, .
\ee
With $\Tr L_{\bf k}^\pm = 8(A_1 + A_2)$ this yields    
\be
{\rm det}(\lambda - L_{\bf k}^\pm) = \lambda^4(\lambda-p_1)^4(\lambda-p_2)^4
\,\, .
\ee
Consequently, both $L_{\bf k}^+$ and $L_{\bf k}^-$ have the eigenvalues 0,
$p_1$, and $p_2$, each with degeneracy 4, which confirms the 
corresponding entries in Table \ref{tableA}.

Finally, we consider the projectors ${\cal P}_{{\bf k},r}^\pm$ corresponding
to the eigenvalues $\lambda$. With 
\be
{\cal P}_{{\bf k},1/2}^\pm=\frac{L_{\bf k}^\pm(L_{\bf k}^\pm-\lambda_{2/1})}
{\lambda_{1/2}(\lambda_{1/2} - \lambda_{2/1})} \,\, , \qquad 
{\cal P}_{{\bf k},3}^\pm = {\bf 1} - {\cal P}_{{\bf k},1}^\pm - 
{\cal P}_{{\bf k},2}^\pm \,\, ,
\ee
one obtains
\begin{subequations} \label{pA}
\bea 
{\cal P}_{{\bf k},1}^\pm &=& \frac{1}{2} J_3^2 \left(1\pm\frac{i}
{\sqrt{A_1A_2-B^2}}\,Z\right) \,\, , \\
{\cal P}_{{\bf k},2}^\pm &=& \frac{1}{2} J_3^2 \left(1\mp\frac{i}
{\sqrt{A_1A_2-B^2}}\,Z\right) \,\, , \\
{\cal P}_{{\bf k},3}^\pm &=& 1 -J_3^2 \,\, .
\eea
\end{subequations}
Since $\Tr Z = \Tr (\Lambda_{\bf k}^e Z) = 0$, Eq.\ (\ref{traces}) is  
obviously fulfilled also in the A phase.

\section{Traces and angular integrals for the transverse phases}
\label{transversephases}

In this appendix, we present the technical details of the solution of the
gap equation for the transverse polar, planar, and A phases. The transverse 
CSL phase is trivial with respect to the angular structure and has already 
been discussed in Ref.\ \cite{schmitt}. 
We compute the color and
Dirac traces and perform the $\uk$-integration of Eq.\ (\ref{pullout}) in
order to determine the coefficients $a_s$ and $\eta^{\ell,t}$ introduced in 
Eqs.\ (\ref{eta}). 
These coefficients yield the constant $d$, cf.\ Eq.\ (\ref{d}).  

Let us define the following traces over Dirac space,
\begin{subequations} \label{defQR}
\bea
{\cal Q}_{\mu\nu}^{mn} &\equiv& {\rm Tr}\left[\g_\mu\g_{\perp,m}(\uq)\,\Lambda_{\bf q}^-\,
\g_\nu\g_{\perp,n}(\uk)\,\Lambda_{\bf k}^+\right] \,\, , \\
{\cal R}_{\mu\nu}^{mn} &\equiv& {\rm Tr}\left[\g_\mu\g_{\perp,m}(\uq)\,\g_0\g_5
\vg\cdot\uq\,\Lambda_{\bf q}^-\,\g_\nu\g_{\perp,n}(\uk)\,\Lambda_{\bf k}^+\right] \,\, .
\eea
\end{subequations}
We shall make use of the following results,
\begin{subequations} \label{dirachelp}
\bea
{\cal Q}_{00}^{mn}
&=& -\d_{mn}(1+\uq\cdot\uk) + \hk_m\hk_n + \hq_m\hq_n-\hq_m\hk_n\,\uq\cdot\uk
+\hq_n\hk_m \,\, , \label{dirachelp1}\\
(\d^{ij} -\hp^i\hp^j)\,{\cal Q}_{ij}^{mn}&=&
2\,\left[-\hp_m\hp_n + \hq_n\hk_m - \uq\cdot\uk \,(\d_{mn}-\hp_m\hp_n)
+\hp_m(\hk_n-\hq_n)\,\up\cdot\uk \right. \nonumber \\ 
&& \left. -\hp_n(\hk_m-\hq_m)\,\up\cdot\uq
+(\d_{mn}-\hq_m\hk_n)\,\up\cdot\uq\,\up\cdot\uk \right] \,\, , \label{dirachelp2}\\
{\cal R}_{00}^{12} - {\cal R}_{00}^{21} &=& i(\hq_3+\hk_3)(1+\uq\cdot\uk) \,\, , \\
(\delta^{ij}-\hp^i\hp^j)({\cal R}_{ij}^{12} - {\cal R}_{ij}^{21})
&=& -2i\,\left\{[1-(\up\cdot\uk)^2]\,\hq_3+[1-(\up\cdot\uq)^2]\,\hk_3-(1-\uq\cdot\uk)
(\up\cdot\uk + \up\cdot\uq)\,\hp_3\right\} \,\, .
\eea 
\end{subequations}

\subsection{Transverse polar phase}

Using the matrix ${\cal M}_{\bf k}$ from Table \ref{tablepolar} with
$(\a,\b) = (0,1)$, we find 
\be
{\rm Tr}\left[{\cal M}_{\bf k}{\cal M}_{\bf k}^\dag\Lambda_{\bf k}^+\right]
=4\,(1-\hk_3^2) \,\, ,
\ee
and thus
\be \label{T00polar}
{\cal T}_{00}^1({\bf k},{\bf q}) = \frac{1}{3\,(1-\hk_3^2)}
\left[1+\uq\cdot\uk-\hq_3\hk_3\,(1-\uq\cdot\uk)-\hk_3^2-\hq_3^2\right] \,\, ,
\ee
where the color trace ${\rm Tr}[T_a^TJ_3 T_a J_3]=-4/3$ has been used.
Analogously, we find
\be \label{Ttpolar}
{\cal T}_t^1({\bf k},{\bf q}) = \frac{2}{3\,(1-\hk_3^2)}\,\left[-\hq_3\hk_3
+\uq\cdot\uk+\hp_3^2\,(1-\uq\cdot\uk) - \hp_3\,(\hq_3-\hk_3)\,
(\up\cdot\uq - \up\cdot\uk) -(1-\hq_3\hk_3)\,\up\cdot\uq\,\up\cdot\uk \right]
\,\, .
\ee
In order to obtain these results, 
Eqs.\ (\ref{dirachelp1}) and (\ref{dirachelp2}) with $m=n=3$ have been employed.
As expected, the ungapped excitation branch does not contribute to the 
gap equation, ${\cal T}_{00}^2({\bf k},{\bf q})={\cal T}_t^2({\bf k},{\bf q})
=0$, and thus $a_1 = 1$, $a_2 = 0$.
 
\begin{figure}[ht] 
\begin{center}
\includegraphics[width=13cm]{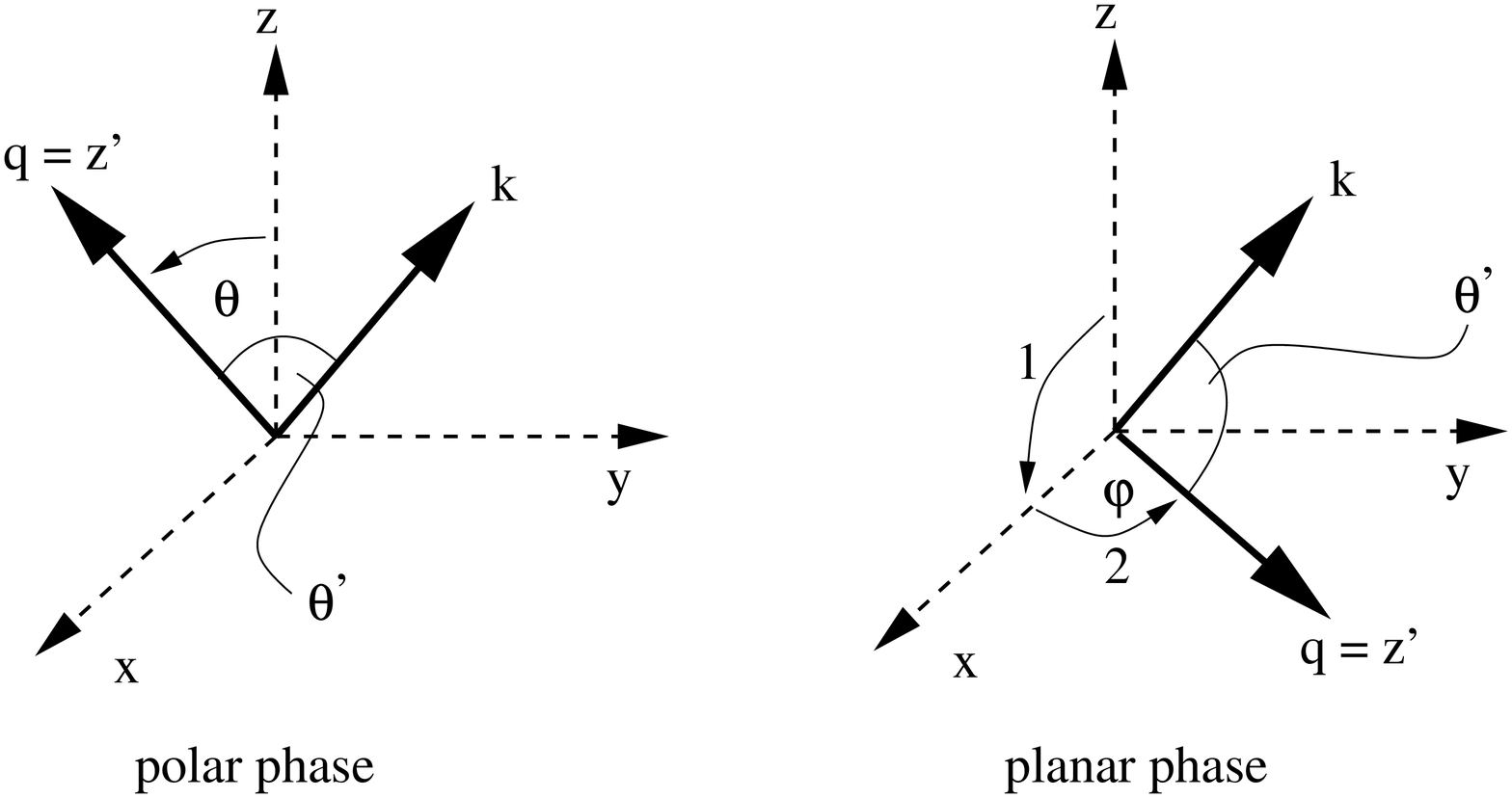}
\vspace{0.5cm}
\caption{Rotation of the coordinate system $(x,y,z) \to (x',y',z')$ 
for the $\uk$-integration in the cases of the transverse polar and 
planar phases.}
\label{figureframes}
\end{center}
\end{figure}

Next, the angular integral $d\Omega_k$ has to be performed. To this end,
we use the following frame (cf.\ left diagram in Fig.\ \ref{figureframes}): 
The order parameter in the polar phase picks
a special direction in real space. By convention, we choose this direction 
to be parallel to the $z$-axis. Now there is another fixed direction, $\uq$,
which we can choose to be in the $xz$-plane, $\uq = 
(\sin\theta,0,\cos\theta)$, where $\theta$ is the polar angle. In order
to perform the $d\Omega_k$ integral, we use a frame with $z'$-axis parallel
to $\uq$. The original frame can be transformed into the new one with 
a rotation $R(\theta)$ around the $y$-axis,
\be
R(\theta) = \left(\begin{array}{ccc} \cos\theta & 0 & -\sin\theta \\
0 & 1 & 0 \\ \sin\theta & 0 & \cos\theta \end{array}\right)\,\, .
\ee
In the new frame, we have $\uq' = (0,0,1)$ and $\uk' = 
(\sin\theta'\cos\varphi',\sin\theta'\sin\varphi',\cos\theta')$. Therefore, 
before performing the $d\Omega_k$ integral, we write $\uk$ as 
\be
\uk = R^{-1}(\theta)\,\uk' = \left(\begin{array}{c} \cos\theta\sin\theta'
\cos\varphi' + \sin\theta\cos\theta' \\ \sin\theta'\sin\varphi' \\
-\sin\theta\sin\theta'\cos\varphi' + \cos\theta\cos\theta'
\end{array}\right)\,\, . 
\ee

This new frame is particularly convenient, since the new polar angle $\theta'$ is the angle between $\uq$ and $\uk$, $\cos\theta' = \uq\cdot\uk$. Therefore,
the integration over $\theta'$ can be transformed into an integral over
$p$ via $\cos\theta'=(k^2+q^2-p^2)/(2kq)$, where $p$ is the modulus of 
${\bf p} = {\bf k} -{\bf q}$. Note that the function $F_\ell$ on the 
right-hand side of Eq.\ (\ref{pullout}) does not depend on the azimuthal angle
$\varphi'$ but only on $p$, and the function $F_t$ only depends on 
$p$ and $\theta$, but not on $\varphi'$ either (neglecting the $\uk$ dependence
in $\e_{{\bf k},r}$). 
Consequently, we have to multiply ${\cal T}_{00}^1$ and ${\cal T}_t^1$, 
given in Eqs.\ (\ref{T00polar}) and (\ref{Ttpolar}), respectively, with
the factor $\ell_k/\ell_q = (1-\hk_3^2)/(1-\hq_3^2)$,
write the results in terms of the new coordinates and integrate over 
$\varphi'$. After setting
$k\simeq q\simeq\mu$, which is permissible to subleading order, the result
yields the coefficients $\eta$ in Eq.\ (\ref{eta}). Then, the result of the 
$p$ integral obtained by Eq.\ (\ref{d}), following Ref.\ \cite{schmitt}.

The explicit results of this procedure are as follows. 
With 
\begin{subequations}
\bea
\frac{1}{2\pi} \int_0^{2\pi} d\varphi'\, \hk_i &=& \hq_i\,\uq\cdot\uk 
\,\, , \label{aziint} \\
\frac{1}{2\pi} \int_0^{2\pi} d\varphi'\, \hk_i^2 &=& \hq_i^2\,(\uq\cdot\uk)^2
+\frac{1}{2}\,(1-\hq_i^2)\,\left[1-(\uq\cdot\uk)^2\right] \,\, ,
\eea
\end{subequations}
we obtain 
\be
\frac{1}{2\pi} \int_0^{2\pi} d\varphi'\, \frac{\ell_k}{\ell_q}\,
{\cal T}_{00}^1({\bf k},{\bf q})
\simeq\frac{1}{2\pi} \int_0^{2\pi} d\varphi'\, \frac{\ell_k}{\ell_q}\,
{\cal T}_t^1({\bf k},{\bf q})
\simeq \frac{2}{3} - \frac{1}{3}\frac{p^2}{\mu^2} + 
\frac{1}{24}\frac{p^4}{\mu^4} \,\, ,
\ee
where the approximation $k\simeq q\simeq\mu$ has been implemented via the 
replacements $\uq\cdot\uk\to 1-p^2/(2\mu^2)$, $\up\cdot\uk = -\up\cdot\uq
\to p/(2\mu)$. Now, we immediately compute $d = 9/2$, as listed in Table
\ref{tablepolar}.

\subsection{Transverse planar phase}

In this case, we find from Table \ref{tableplanar}
\be \label{denominatorplanar}
{\rm Tr}\left[{\cal M}_{\bf k}{\cal M}_{\bf k}^\dag\Lambda_{\bf k}^+\right]
=4\,(1+\hk_3^2) \,\, .
\ee
Furthermore,
\be \label{projectorplanar}
{\cal P}_{{\bf q},1}^+ = \frac{1}{1+\hq_3^2}\left[J_1^2(1-\hq_1^2)+J_2^2(1-\hq_2^2)
-\{J_1,J_2\}\,\hq_1\hq_2+J_3\hq_3\,\g_0\g_5\vg\cdot\uq\right] \,\, ,
\ee
where we have applied the identity
\be \label{gamma5}
\g_{\perp,1}(\uq)\g_{\perp,2}(\uq)-\hq_1\hq_2 = i\hq_3\,\g_0\g_5\vg\cdot\uq \,\, .
\ee
With the help of Eqs.\ (\ref{denominatorplanar}), (\ref{projectorplanar}) and the
definitions (\ref{defT}), (\ref{Ttrans}), we compute the quantities 
${\cal T}^s_{00}({\bf k},{\bf q})$ and ${\cal T}^s_t({\bf k},{\bf q})$. 
As in the polar phase, we find ${\cal T}^2_{00}({\bf k},{\bf q}) = 
{\cal T}^2_t({\bf k},{\bf q})=0$, hence $a_1=1$, $a_2=0$.
For the gapped branch, $s=1$, one obtains after taking the trace over color space,
\be
{\cal T}^1_{\mu\nu}({\bf k},{\bf q}) = - \frac{(\hq_1^2-\hq_3^2-2)\,
{\cal Q}_{\mu\nu}^{11} + (\hq_2^2-\hq_3^2-2)\,{\cal Q}_{\mu\nu}^{22} + 
\hq_1\hq_2\left({\cal Q}_{\mu\nu}^{12} + {\cal Q}_{\mu\nu}^{21}\right) 
+ \;i\hq_3\left({\cal R}_{\mu\nu}^{12} -
{\cal R}_{\mu\nu}^{21}\right)}
{6(1+\hk_3^2)(1+\hq_3^2)}
 \,\, ,
\ee
where the tensors ${\cal Q}$ and ${\cal R}$ are defined in Eqs.\ (\ref{defQR}).
In order to perform the traces over Dirac space, one makes use of the identities 
(\ref{dirachelp}).
We do not present the explicit results for ${\cal T}^1_{00}({\bf k},{\bf q})$ and 
${\cal T}^1_t({\bf k},{\bf q})$ since they are too lengthy. 
In order to perform the angular integration, we proceed in a similar 
way as discussed above for the transverse polar phase. The difference is that
in the planar phase the order parameter does not point into a special direction, 
but is located in the $xy$-plane. Without loss of generality, we can assume
$\uq$ to be also in the $xy$-plane, $\uq = (\cos\varphi,\sin\varphi,0)$
(cf.\ right diagram in Fig.\ \ref{figureframes}). Note that this choice, in particular 
$\hq_3=0$, immediately shows that the term proportional to $\g_5$ on the right-hand
side of Eq.\ (\ref{projectorplanar}) yields no contribution to the integral.  
Again, we rotate the frame such that the $z'$-axis is parallel to 
$\uq$. This can be done with two successive rotations $R_1$ and $R_2(\varphi)$:
First, we rotate the original frame by $\pi/2$ around the $y$-axis, 
\be  
R_1 = \left(\begin{array}{ccc} 0 & 0 & -1 \\ 0 & 1 & 0 \\ 1 & 0 & 0 
\end{array}\right) \,\, .
\ee
Then, we rotate by $\varphi$ around the new $x'$-axis,
\be  
R_2(\varphi) = \left(\begin{array}{ccc} 1 & 0 & 0 \\ 0 & \cos\varphi & 
-\sin\varphi \\ 0 & \sin\varphi & \cos\varphi 
\end{array}\right) \,\, .
\ee
As above, in the new frame, $\uq' = (0,0,1)$ and $\uk' = 
(\sin\theta'\cos\varphi',\sin\theta'\sin\varphi',\cos\theta')$, and hence 
\be
\uk =  R_1^{-1}\,R_2^{-1}(\varphi)\,\uk' = \left(\begin{array}{c} 
-\sin\theta'\sin\varphi'\sin\varphi + \cos\theta'\cos\varphi \\
\sin\theta'\sin\varphi'\cos\varphi + \cos\theta'\sin\varphi \\
-\sin\theta'\cos\varphi' \end{array}\right)\,\, . 
\ee
With $\ell_k/\ell_q =(1+\hk_3^2)/(1+\hq_3^2)$ and the approximation 
$k\simeq q\simeq\mu$, we obtain 
\be
\frac{1}{2\pi} \int_0^{2\pi} d\varphi'\, \frac{\ell_k}{\ell_q}\,
{\cal T}_{00}^1({\bf k},{\bf q})
\simeq\frac{1}{2\pi} \int_0^{2\pi} d\varphi'\, \frac{\ell_k}{\ell_q}\,
{\cal T}_t^1({\bf k},{\bf q})
\simeq \frac{2}{3} - \frac{1}{3}\frac{p^2}{\mu^2} + 
\frac{1}{24}\frac{p^4}{\mu^4} \,\, .
\ee
From these results we conclude $d=9/2$.  

\subsection{Transverse A phase}

In the transverse A phase, we have 
\be
{\rm Tr}\left[{\cal M}_{\bf k}{\cal M}_{\bf k}^\dag\Lambda_{\bf k}^+\right]
=4\,(1+\hk_3^2) \,\, ,
\ee
The projectors corresponding to
the gapped excitations, ${\cal P}_{{\bf q},1/2}^+$ in Eqs.\ (\ref{pA}),
are 
\be \label{PtransverseA}
{\cal P}_{{\bf q},1/2}^+ = \frac{1}{2}\,J_3^2\,\left\{ 1 \mp \frac{i}{|\hq_3|}
\,\left[\g_{\perp,1}(\uq)\,\g_{\perp,2}(\uq) -\hq_1\hq_2\right]\right\} 
= \frac{1}{2}\,J_3^2\,\left( 1 \pm \frac{\hq_3}{|\hq_3|}\g_0\g_5\vg\cdot\uq\right) \,\, ,
\ee
where Eq.\ (\ref{gamma5}) has been used. Inserting this projector and the respective 
quantities from Table \ref{tableA} into Eq.\ (\ref{defT}), one obtains after
taking the trace over color space,
\be \label{TmunuA}
{\cal T}_{\mu\nu}^{1/2}({\bf k},{\bf q}) = \frac{{\rm Tr}\left[\g_\mu\,[\g_{\perp,1}(\uq) 
+ i\g_{\perp,2}(\uq)]\,(1\pm \frac{\hq_3}{|\hq_3|}\g_0\g_5\vg\cdot\uq)\,\Lambda_{\bf q}^-
\,\g_\nu\,[\g_{\perp,1}(\uk) - i\g_{\perp,2}(\uk)]\,\Lambda_{\bf k}^+\right] }
{6(1+\hk_3^2)} \,\, ,
\ee
and ${\cal T}^3_{\mu\nu}({\bf k},{\bf q})=0$.
Again, the results for the quantities ${\cal T}^{1,2}_{00}({\bf k},{\bf q})$ and 
${\cal T}^{1,2}_t({\bf k},{\bf q})$, corresponding to the gapped branches, are 
complicated functions of $\uk$ and $\uq$, and we do not present their 
explicit forms.

As in the transverse planar phase, the order parameter for the transverse A phase 
points into a direction 
perpendicular to the $z$-axis. However, in order to perform the angular integral over 
$\varphi'$, we cannot make use of the same choice of the frame, because
the assumption of $\uq$ being in the $xy$-plane, i.e., $\hq_3=0$, would lead to a 
division by zero according to Eq.\ (\ref{TmunuA}). 
Instead, we allow for the most general form, 
$\uq = (\sin\theta\cos\varphi,\sin\theta\sin\varphi, \cos\theta)$. 
In this case, the rotation of the original frame into the one 
with $z'$-axis parallel to $\uq$ is given by 
\be \label{generalrotation}
R(\theta,\varphi)=\left(\begin{array}{ccc} \cos\theta\,\cos\varphi & 
\cos\theta\,\sin\varphi & -\sin\theta \\ -\sin\varphi & \cos\varphi & 0 \\
\sin\theta\,\cos\varphi & \sin\theta\,\sin\varphi & \cos\theta \end{array}
\right) \,\, .
\ee
Consequently, the $d\Omega_k$ integral has to be performed with 
\be
\uk = R^{-1}(\theta,\varphi)\,\uk' = 
\left(\begin{array}{c} \cos\varphi\cos\theta
\sin\theta'\cos\varphi'-\sin\varphi\sin\theta'\sin\varphi'+\cos\varphi
\sin\theta\cos\theta' \\ \sin\varphi\cos\theta\sin\theta'\cos\varphi' + 
\cos\varphi\sin\theta'\sin\varphi' + \sin\varphi\sin\theta\cos\theta' \\
-\sin\theta\sin\theta'\cos\varphi' + \cos\theta\cos\theta' \end{array}
\right) \,\, ,
\ee
where $\uk'=(\sin\theta'\cos\varphi',\sin\theta'\sin\varphi',\cos\theta')$ has 
been employed. 

With $k\simeq q\simeq\mu$ one obtains
\be
\frac{1}{2\pi} \int_0^{2\pi} d\varphi'\, \frac{\ell_k}{\ell_q}\,
{\cal T}_{00}^{1/2}({\bf k},{\bf q})
\simeq\frac{1}{2\pi} \int_0^{2\pi} d\varphi'\, \frac{\ell_k}{\ell_q}\,
{\cal T}_t^{1/2}({\bf k},{\bf q})
\simeq \frac{1}{2}\left(\frac{2}{3} - \frac{1}{3}\frac{p^2}{\mu^2} + 
\frac{1}{24}\frac{p^4}{\mu^4}\right)\left(1\pm 2\frac{|\cos\theta|}{1+\cos^2\theta}
\right) \,\, .
\ee
The term $\pm\,2\,|\cos\theta|\,/\,(1+\cos^2\theta)$ on the right-hand side of this 
equation gives rise to a difference between the quantities 
${\cal T}_{00,t}^{1}({\bf k},{\bf q})$ and ${\cal T}_{00,t}^2({\bf k},{\bf q})$.
The origin of this difference is the term proportional to $\g_5$ on the right-hand side
of Eq.\ (\ref{PtransverseA}). As a consequence, one cannot choose the coefficients 
$a_s$ such that they are constants, cf.\ Eqs.\ (\ref{eta}). The transverse A phase is the
only phase we consider, in which these coefficients depend on $\theta$, the angle 
between $\uq$ and the $z$-axis. We find
\be
a_1 = \frac{1}{2} + \frac{|\cos\theta|}{1+\cos^2\theta} \,\, , \qquad 
a_2 = \frac{1}{2} - \frac{|\cos\theta|}{1+\cos^2\theta} \,\, , \qquad 
a_3 = 0 \,\, .
\ee
However, the constant $d$ is identical to all other transverse phases, $d=9/2$.

\section{Proving $\overline{d}=6$ for arbitrary longitudinal gaps}
\label{AppB}

In this appendix, we prove $\overline{d}=6$ for any order parameter 
(= for any $3\times 3$ matrix) $\Delta$ in the case of a longitudinal gap.  
The longitudinal case is particularly simple since the Dirac structure of
the matrix ${\cal M}_{\bf k}$ is trivial. 

Let us start with the matrix 
\be
{\cal M}_{\bf k} = \vv_{\bf k}\cdot \vJ \,\, ,
\ee
where $\vv_{\bf k}=(v_{{\bf k},1},v_{{\bf k},2},v_{{\bf k},3})$ is a 3-vector
with
\be
v_{{\bf k},i} \equiv \sum_{j=1}^3 \Delta_{ij}\hat{k}_j \,\, ,
\qquad i = 1,2,3 \,\, .
\ee
Then,
\be
(L_{\bf k}^+)_{ij} = v_{\bf k}^2\,\d_{ij} - v^*_{{\bf k},j}
\,v_{{\bf k},i} \,\, ,
\ee
where $v_{\bf k}^2\equiv\vv_{\bf k}^*\cdot\vv_{\bf k}$. Now, with 
$(L_{\bf k}^+)^2=v_{\bf k}^2\,L_{\bf k}^+$  
and $\Tr \,L_{\bf k}^+ = 8\,v^2_{\bf k}$ the eigenvalues of $L_{\bf k}^+$ are 
easily found making use of the method shown in Appendix 
\ref{AppA}. One obtains 
\be
\lambda_{k,1} = v_{\bf k}^2 \qquad (\mbox{8-fold}) \,\, , 
\qquad \lambda_{k,2} = 0 \qquad (\mbox{4-fold}) \,\, . 
\ee 
Consequently, the fact that there is one gapped branch with degeneracy 8
and one ungapped branch with degeneracy 4 is completely general, i.e., it 
is true for any order parameter in the longitudinal case.  
The corresponding projectors are given by
\be
({\cal P}_{{\bf k},1}^+)_{ij}  = \d_{ij} - 
\frac{v_{{\bf k},j}^* v_{{\bf k},i}}{v^2_{\bf k}} \,\, , \qquad
({\cal P}_{{\bf k},2}^+)_{ij}  = \frac{v_{{\bf k},j}^* v_{{\bf k},i}}
{v_{\bf k}^2} 
\,\, .
\ee
This leads to 
\be
{\cal T}_{00}^1({\bf k},{\bf q})=\frac{1}{3}\,
\frac{\vv_{\bf q}\cdot\vv_{\bf k}^*}{v_{\bf k}^2}\,(1+\uq\cdot\uk)\,\, ,
\qquad
{\cal T}_t^{1}({\bf k},{\bf q})=\frac{2}{3}\,
\frac{\vv_{\bf q}\cdot\vv_{\bf k}^*}{v_{\bf k}^2}\,(1-\up\cdot\uq\,
\up\cdot\uk) \,\, ,
\ee
and ${\cal T}_{00}^2({\bf k},{\bf q})={\cal T}_t^2({\bf k},{\bf q})=0$, 
hence $a_1=1$, $a_2=0$. The angular integration is done 
as explained for the transverse A phase in the previous 
appendix, i.e., the rotation given by Eq.\ (\ref{generalrotation}) and the 
respective expressions for the vectors $\uk$ and $\uq$ are used.

With $\ell_k/\ell_q=\lambda_{k,1}/\lambda_{q,1}= v_{\bf k}^2/v_{\bf q}^2$ 
we obtain
\be
\frac{1}{2\pi}\int_0^{2\pi}d\varphi'\,\frac{\ell_k}{\ell_q}
\frac{\vv_{\bf q}\cdot\vv_{\bf k}^*}{v_{\bf k}^2} = \uq\cdot\uk \,\, ,
\ee
where Eq.\ (\ref{aziint}) has been used. Therefore, we find with 
$k\simeq q\simeq \mu$,
\be
\frac{1}{2\pi}\int_0^{2\pi}d\varphi'\, \frac{\ell_k}{\ell_q} \,{\cal T}_{00}^1
({\bf k},{\bf q})\simeq
\frac{1}{2\pi}\int_0^{2\pi}d\varphi'\,\frac{\ell_k}{\ell_q} \,{\cal T}_t^1
({\bf k},{\bf q})\simeq \frac{2}{3} - \frac{1}{2}\frac{p^2}{\mu^2} + \frac{1}{12}\frac{p^4}{\mu^4}
\,\, ,
\ee
which, making use of the definitions (\ref{eta}) and (\ref{d}), 
proves the universal result $d=6$ for longitudinal gaps. Since 
$d=\overline{d}$ when $d$ is constant, cf.\ definition (\ref{zeta}), 
we conclude $\overline{d}=6$.

\end{appendix}


\begin{thebibliography}{99}

\bibitem{bcs} 
J.\ Bardeen, L.N.\ Cooper, and J.R.\ Schrieffer, 
Phys.\ Rev.\ {\bf 108}, 1175 (1957).

\bibitem{bailin} 
D.\ Bailin and A.\ Love, 
Phys.\ Rep.\ {\bf 107}, 325 (1984).

\bibitem{alford} 
M.\ Alford, K.\ Rajagopal, and F.\ Wilczek, 
Nucl.\ Phys.\ {\bf B537}, 443 (1999). 

\bibitem{alford2}
M.G.\ Alford, J.A.\ Bowers, and K.\ Rajagopal, Phys.\ Rev.\ D {\bf 63}, 
074016 (2001).

\bibitem{loff}
A.I.\ Larkin and Y.N.\ Ovchinnikov, Zh.\ Eksp.\ Teor.\ Fiz.\ {\bf 47},
1136 (1964); P.\ Fulde and R.A.\ Ferrell, Phys.\ Rev.\ {\bf 135}, A550 (1964).

\bibitem{shovkovy}
I.\ Shovkovy and M.\ Huang, Phys.\ Lett.\ B {\bf 564}, 205 (2003);
Nucl.\ Phys.\ {\bf A729}, 835 (2003).

\bibitem{gCFL}
M.\ Alford, C.\ Kouvaris, and K.\ Rajagopal, Phys.\ Rev.\ Lett.\ {\bf 92},
222001 (2004);
S.B.\ R\"uster, I.A.\ Shovkovy, and D.H.\ Rischke, Nucl.\ Phys.\ {\bf A743},
127 (2004);
K.\ Fukushima, C.\ Kouvaris, and K.\ Rajagopal, hep-ph/0408322.

\bibitem{kaon}
P.F.\ Bedaque and T.\ Sch\"afer, Nucl.\ Phys.\ {\bf A697}, 802 (2002);
A.\ Kryjevski, D.B.\ Kaplan, and T.\ Sch\"afer, hep-ph/0404290;
A.\ Kryjevski and D.\ Yamada, hep-ph/0407350.

\bibitem{muether}
H.\ M\"uther and A.\ Sedrakian, Phys.\ Rev.\ D {\bf 67}, 085024 (2003).

\bibitem{oneflavor} 
M.\ Iwasaki and T.\ Iwado, 
\PLB{350}{163}{1995};
R.D.\ Pisarski and D.H.\ Rischke,
\PRD{61}{051501}{2000};
M.G.\ Alford, J.A.\ Bowers, J.M.\ Cheyne, and G.A.\ Cowan,
\PRD{67}{054018}{2003};
M.\ Buballa, J.\ Ho\v{s}ek, and M.\ Oertel,
\PRL{90}{182002}{2003}.

\bibitem{rischke}
R.D.\ Pisarski and D.H.\ Rischke,
\PRD{61}{074017}{2000};

\bibitem{schaefer} 
T.\ Sch\"afer, 
Phys.\ Rev.\ D {\bf 62}, 094007 (2000).

\bibitem{leggett}
A.J.\ Leggett, Rev.\ Mod.\ Phys.\ {\bf 47}, 331 (1975).

\bibitem{vollhardt} 
D.\ Vollhardt and P.\ W\"olfle, 
{\it The Superfluid Phases of Helium 3} (Taylor \& Francis, London, 1990).  

\bibitem{QCDgapeq}
T.\ Sch\"afer and F.\ Wilczek,
\PRD{60}{114003}{1999};
D.K.\ Hong, V.A.\ Miransky, I.A.\ Shovkovy, and L.C.R.\ Wijewardhana,
\PRD{61}{056001}{2000}; Erratum {\it ibid.} D {\bf 62}, 059903 (2000).

\bibitem{son}
D.T.\ Son, \PRD{59}{094019}{1999}.

\bibitem{ren} 
W.E.\ Brown, J.T.\ Liu, and  H.-C.\ Ren, 
\PRD{61}{114012}{2000}; 
\ibid{62}{054013}{2000}; 
\ibid{62}{054016}{2000}. 

\bibitem{wang}
Q.\ Wang and D.H.\ Rischke, \PRD{65}{054005}{2002}.

\bibitem{schmitt}
A.\ Schmitt, Q.\ Wang, and D.H.\ Rischke, \PRD{66}{114010}{2002}.

\bibitem{michel}
L.\ Michel, Rev.\ Mod.\ Phys.\ {\bf 52}, 617 (1980).

\bibitem{schmitt2}
A.\ Schmitt, Q.\ Wang, and D.H.\ Rischke, \PRL{91}{242301}{2003}.

\bibitem{schmitt3}
A.\ Schmitt, Q.\ Wang, and D.H.\ Rischke, \PRD{69}{094017}{2004}.

\bibitem{cjt}
J.M.\ Cornwall, R.\ Jackiw, and E.\ Tomboulis, \PRD{10}{2428}{1974}.

\bibitem{review}
D.H.\ Rischke, Prog.\ Part.\ Nucl.\ Phys.\ {\bf 52}, 197 (2004). 

\bibitem{abuki}
H.\ Abuki, Prog.\ Theor.\ Phys.\ {\bf 110}, 937 (2003).

\bibitem{ruester}
S.B.\ R\"uster and D.H.\ Rischke, \PRD{69}{045011}{2004}.

\bibitem{manuel}
C.\ Manuel, \PRD{62}{114008}{2000}.

\bibitem{thesis}
A.\ Schmitt, {\em Ph.D.\ thesis}, nucl-th/0405076.


\bibitem{rischke2}
D.H.\ Rischke, \PRD{64}{094003}{2001}.


\bibitem{miransky}
V.A.\ Miransky, I.A.\ Shovkovy, and L.C.R.\ Wijewardhana, \PLB{468}{270}{1999};
I.A.\ Shovkovy and P.J.\ Ellis, Phys.\ Rev.\ C {\bf 66}, 015802 (2002).




\bibitem{meissner3}
D.H.\ Rischke, \PRD{62}{054017}{2000}.

\bibitem{neutrality}
A.\ Gerhold and A.\ Rebhan, \PRD{68}{011502}{2003};
D.D.\ Dietrich and D.H.\ Rischke, Prog.\ Part.\ Nucl.\ Phys.\ {\bf 53}, 305
(2004).

\bibitem{tsuei}
C.C.\ Tsuei and J.R.\ Kirtley, Rev.\ Mod.\ Phys.\ {\bf 72}, 969 (2000);
K.\ Izawa, K.\ Kamata, Y.\ Nakajima, Y.\ Matsuda, T.\ Watanabe, M.\ Nohara, 
H.\ Takagi, P.\ Thalmeier, and K.\ Maki, \PRL{89}{137006}{2002}.

\bibitem{weber}
F.\ Weber, {\em Pulsars as Astrophysical Laboratories for Nuclear and Particle Physics} 
(IOP Publishing Ltd., Bristol, 1999); 
M.\ Prakash, J.M.\ Lattimer, J.A.\ Pons, A.W.\ Steiner, 
and S.\ Reddy, Lect.\ Notes Phys.\ {\bf  578}, 364 (2001).

\bibitem{hotwater}
M.\ Alford, P.\ Jotwani, C.\ Kouvaris, J.\ Kundu, and K.\ Rajagopal, astro-ph/0411560.

\bibitem{blaschke}
H.\ Grigorian, D.\ Blaschke, and D.\ Voskresensky, astro-ph/0411619.

\end{thebibliography}
\end{document}